\PassOptionsToPackage{sort&compress}{natbib}
\documentclass[1p]{elsarticle} %twocolumn

\usepackage{amsmath,amssymb,amsfonts,xspace,subfigure}
\usepackage{graphicx}
\usepackage[colorlinks=true, allcolors=blue]{hyperref}
\usepackage[nameinlink]{cleveref}
\usepackage[dvipsnames]{xcolor}
\usepackage{url}
\PassOptionsToPackage{
  colorlinks=true, citecolor=blue,
  linkcolor=blue, urlcolor=teal
}{hyperref}
\usepackage{fullpage}
\usepackage{cases}
\usepackage{setspace}
\newcommand{\yc}[1]{{\color{orange} #1}}

\newcommand{\Bepsilon}{{\boldsymbol{\varepsilon}}}
\newcommand{\Bsigma}{{\boldsymbol{\sigma}}}
\newcommand{\Bn}{{\mathbf{n}}}
\newcommand{\Bu}{{\mathbf{u}}}
\newcommand{\Br}{{\mathbf{r}}}
\newcommand{\BT}{{\mathbf{T}}}
\newcommand{\Bk}{{\mathbf{k}}}
\newcommand{\BI}{{\mathbf{I}}}
\newcommand{\dA}{\,\mathrm{d}A}
\newcommand{\ds}{\,\mathrm{d}s}
\newcommand{\tra}{^{\sf T}}
\DeclareMathOperator{\Tr}{tr}

\graphicspath{{./Figures/}}

\journal{EML}

\begin{document}

\begin{frontmatter}

\title{Stress distribution in contractile cell monolayers
%Mapping Intercellular Forces in Contractile Cell Monolayers Using Monolayer Stress Microscopy
%Monolayer Stress Microscopy Reveals Stress Distributions in Contractile Cells}
}

\author[1]{Yucheng Huo\fnref{equal}}
\author[1]{Kexin Guo\fnref{equal}}
\author[1]{Massimo Paradiso}
\author[1,2]{K. Jimmy Hsia\corref{cor1}}

\cortext[cor1]{Corresponding author: kjhsia@ntu.edu.sg (K.J.H.)}
\fntext[equal]{These authors contributed equally to this work.}

\address[1]{School of Mechanical and Aerospace Engineering, Nanyang Technological University, Singapore 639798, Singapore}
\address[2]{School of Chemistry, Chemical Engineering and Biotechnology, Nanyang Technological University, Singapore 637371, Singapore}
% \affiliation[1]{
%     organization={School of Mechanical and Aerospace Engineering, Nanyang Technological University},
%     addressline={50 Nanyang Avenue},
%     postcode={639798},
%     city={Singapore},
%     country={Singapore}
% }

% \affiliation[2]{
%     organization={School of Chemistry, Chemical Engineering and Biotechnology, Nanyang Technological University},
%     addressline={21 Nanyang Link},
%     postcode={637371},
%     city={Singapore},
%     country={Singapore}
% }
% 
% \ead{kjhsia@ntu.edu.sg}

\begin{abstract}
% Collective cell behaviors in tissues are intimately regulated by mechanical forces, yet the quantification of intercellular stresses in contractile monolayers remains insufficiently understood. 
% Here, we reconstruct the internal stress fields of cultured C2C12 myoblast monolayers using monolayer stress microscopy to uncover how local mechanical forces organize collective structures. 
% We find that contractile monolayers maintain an intrinsically tensile state, characterized by positive maximum and negative minimum principal stresses, reflecting the anisotropy of active tension. 
% Stress patterns exhibit striking transitions around topological defects, which coincide with singularities in cell alignment, density, and morphology—revealing a strong coupling between mechanical forces and structural organization. 
% Moreover, tensile stresses are preferentially transmitted along the cell elongation axis and compressive stresses transversely, demonstrating that local stresses guide cell arrangement. 
% This mechanical guidance disappears when cells lose their contractility and shape anisotropy, but reemerges in other contractile systems such as bone marrow–derived mesenchymal stem cells. 
% Together, our work puts forward a quantitative strategy to characterize mechanical anisotropy within active cellular monolayers and establishes a universal principle of force–structure coupling, providing a physical basis for understanding how mechanics governs myogenesis, morphogenesis, and collective organization in living tissues.

Collective behaviors in cellular systems are regulated not only by biochemical signalling pathways but also by intercellular mechanical forces, whose quantification in contractile monolayers remains poorly understood. 
Here, by integrating traction force microscopy and numerical simulations, we reconstruct the stress distribution in C2C12 myoblast monolayers to reveal the roles of local mechanical forces in determining the collective cellular structures. 
We find that contractile monolayers exhibit positive maximum and negative minimum principal stresses, reflecting the intrinsic anisotropy of active tension. 
Distinct stress patterns emerge around topological defects, coinciding with singularities in cell alignment, density, and morphology, indicating a strong coupling between mechanical forces and structural organization. 
Moreover, tensile stresses are preferentially transmitted along the cell elongation axis and compressive stresses transversely, demonstrating that local stress guides cell arrangement. 
This mechanical guidance appears to be universal among contractile systems, as observed also in bone marrow–derived mesenchymal stem cells. 
Together, our work establishes a quantitative framework for characterizing mechanical anisotropy in active cellular monolayers and reveals a general principle of force–structure coupling, providing a physical basis for understanding how mechanics governs myogenesis, morphogenesis, and collective organization in contractile cellular systems.
\end{abstract}


\end{frontmatter}

%----------------------------------------------------
\section{Introduction}

The collective behaviors arising from mechanical interactions in cellular monolayers are fundamental to critical biological processes \cite{li_eml_2021}, including tissue development \cite{fagotto_tissue_2020,guillamat_integer_2022,schoenit_force_2025}, wound healing \cite{ravasio_gap_2015,wei_actin-ring_2020,xu_geometry-mediated_2023}, and cancer metastasis \cite{tang_mechanicalTumorMetastasis_2010,zhang_traction_2019,guan_interfacial_2023}. Unlike non-biological materials, living tissues actively generate forces via intracellular contractility and transmit forces through extracellular adhesion \cite{trepat_physicalforcetransmission_2009,tambe_collective_2011TFM+MSM,serra-picamal_mechanicalwavesinExpansion_2012,
vasquez_force_2016}, creating a complex mechanically coordinated network over multiple cell lengths to regulate cell shape \cite{park_unjamming_2015,atia_geometric_2018,luciano_CellNucleusShapeAdaption_2021}, motion \cite{vedula_emerging_2012,doxzen_guidance_2013,vedula_epithelial_2014} and arrangement \cite{duclos_perfect_2014,balasubramaniam_investigating_2021,coyle_cell_2022,ienaga_geometric_2023}, which is critical to maintaining mechanical homeostasis within the collective \cite{lecuit_ReviewForceTransmission_2011,saw_mechanobiology_2015,Ladoux_mechanobiology_2017,zhang_Extracellular&IntracrllularTFM_2019,feng_ReviewMechano-chemo-biological_2025}.

During the past two decades, numerous methodologies have been developed to quantify the forces of mechanical interaction within cellular tissues. Intercellular forces can be measured directly using techniques such as laser ablation of cell junctions \cite{ratheesh_LaserAblationCentralspindlin_2012,smutny_LaserAblationZebrafish_2015,liang_LaserAblationProtocol_2016}, molecular stress sensors based on Förster resonance energy transfer (FRET) \cite{kong_fret_2005,meng_fret_2012,borghi_FRETe-cadherin_2012PNAS}, or tension-sensitive fluorescent proteins \cite{wang_fluorescence_2008,guo_fluorescence_2014}. However, these approaches typically capture forces only at the level of single junctions or molecules, providing highly localized measurements that are insufficient to reveal the collective stress organization across larger cell populations. To address this limitation, monolayer stress microscopy (MSM) has emerged as a powerful method to assess the intra-monolayer stresses by enforcing mechanical equilibrium based on the information obtained from traction force microscopy (TFM) \cite{saez_ReviewTF_2010, trepat_physicalforcetransmission_2009, style_TFMReview_2014Softmatter}. It computes the spatial distribution of mechanical stresses by solving the equilibrium equations that balance substrate traction as external shear with internal stresses, typically implemented through finite element modeling (FEM) \cite{tambe_collective_2011TFM+MSM, tambe_MSMLimitation_2013,zhang_Extracellular&IntracrllularTFM_2019,balasubramaniam_investigating_2021,guan_interfacial_2023,li_EMLmachine_2024,saraswathibhatla_EMLcoordinated_2021}. Stress distribution patterns within tissues contain critical information to understand the underlying mechanical regulatory principles in cell migration, providing valuable insights into the collective behavior of multicellular structures.

Despite the power of MSM to quantify cellular forces, its application has largely been restricted to epithelial monolayers such as MDCK cells \cite{tambe_collective_2011TFM+MSM,zhang_Extracellular&IntracrllularTFM_2019,balasubramaniam_investigating_2021,sonam_HoleFormation_2023}. This leaves mesenchymal tissues—including myoblasts, fibroblasts, and endothelial cells—relatively unexplored, even though mechanical stress is a key regulator of their alignment \cite{coyle_cell_2022,duclos_perfect_2014,grigola_myoblast_2014,ienaga_geometric_2023}, differentiation, and morphogenesis. C2C12 myoblasts represent a canonical model of such systems. These cells exhibit high actomyosin contractility, driving elongation and pronounced mechanical anisotropy \cite{doss_cell_2020,lv_morphodynamics_2024,stehbens_perspectives_2024,saw_mechanobiology_2015}. As we have demonstrated  \cite{huo_myoblast_2025}, intercellular adhesion is stronger along the longitudinal axis of cell–cell contacts, reinforcing this anisotropy. Coordinately, strong directional contractility and adhesive anisotropy promote the spontaneous self-organization of myoblasts into large-scale nematic order, a process that is not only fundamental to active matter physics but also highly relevant to muscle regeneration in tissue engineering \cite{le_grand_skeletal_2007,bentzinger_building_2012}. It is therefore critical to quantify how these anisotropic physical interactions at the cellular level manifest as tissue-scale stress patterns. However, this issue remains largely unanswered, motivating the extension of MSM beyond epithelial contexts.

Here, we apply MSM to evaluate the stress distribution in nematically ordered cell monolayers, including C2C12 and BMSC cells. We find these contractile tissues exhibit a state of anisotropic tension, with positive maximum and negative minimum principal stresses, consistent with our prior findings of stronger longitudinal adhesion \cite{huo_myoblast_2025}. This stress anisotropy is influenced by tissue architecture, e.g.,topological defects play the role of singularity in inducing asymmetric stress patterns and transitional cellular geometry. We then investigate the coupling between this stress state and cellular organization, and demonstrate that the maximum principal stress aligns with the cellular director in ordered states but becomes less aligned in disordered states. Crucially, this coupling between cellular orientation and stress anisotropy is a generalizable principle across other contractile cell types. Together, these findings establish a direct mechanistic link from subcellular contractility and adhesion through cellular anisotropy to tissue-scale stress patterns. This work provides a framework for evaluating mechanical stress in anisotropic tissues and advances the fundamental biophysics of active nematic systems.

\begin{figure}[tbh]
    \centering
    \includegraphics[width=0.9\linewidth]{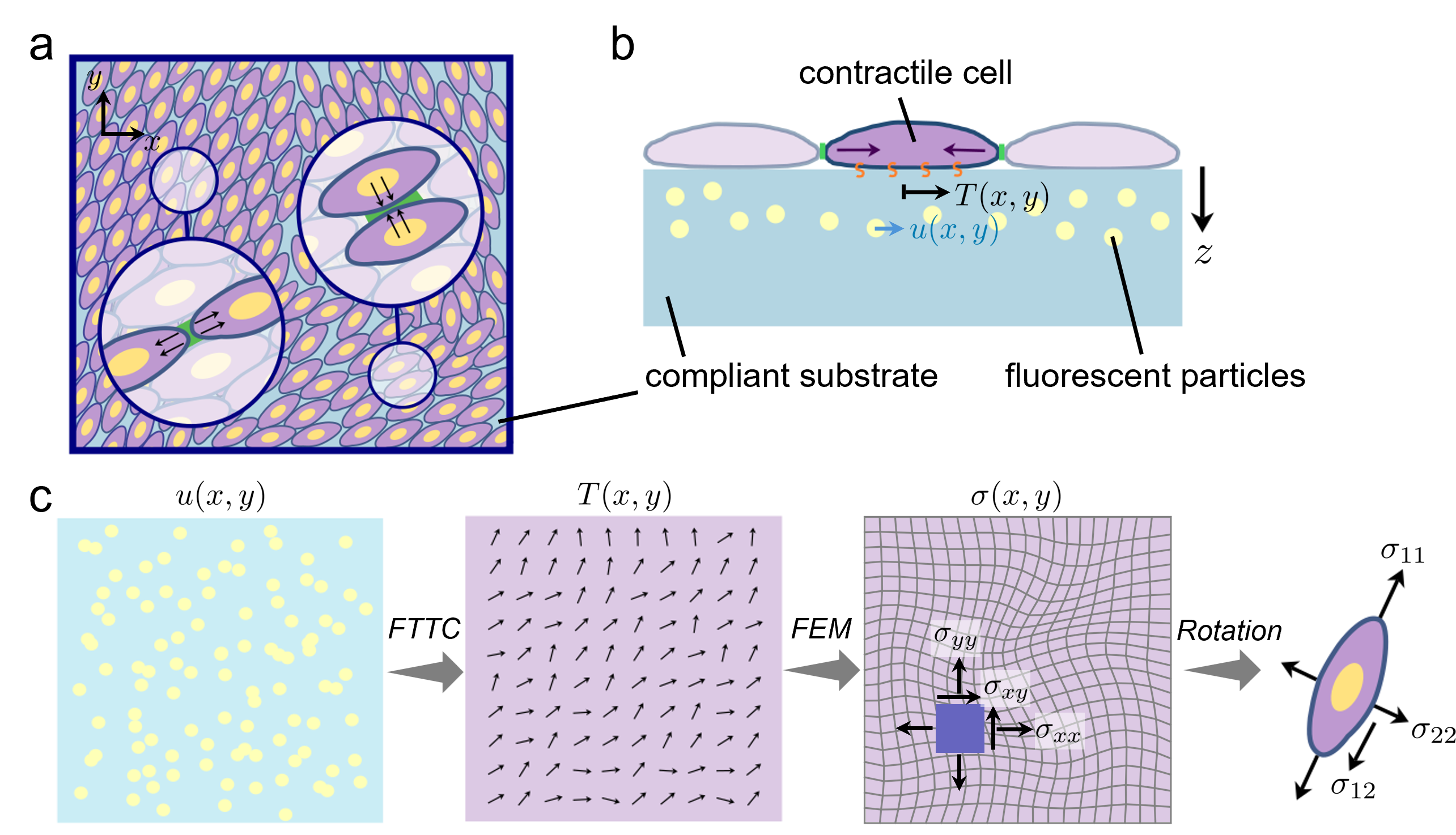}
    \caption{\textbf{Schematic overview of the inverse problem inferring internal stress in a cell monolayer from measured traction forces.} 
    \textbf{a}. Schematic in top-down view of a confluent monolayer of muscle cells on a soft substrate. Intercellular interactions along the longitudinal and transverse directions of the cell. Arrows indicate internal stress between cells. Green bars highlight two types of cell-cell junctions responsible for different interactions.    
    \textbf{b}. Side-view of a cell attached to the soft substrate. Traction forces, \textbf{T}, are exerted to the soft substrate, inducing its deformation and displacement of the fluorescent beads, \textbf{u}. The substrate is assumed to be semi-infinite, extending in z direction.
    \textbf{c}. Methodological workflow. (i) Displacement field of substrate. PIV analysis of fluorescent beads gives the displacement vector fields. (ii) Traction field (hypothetical) is obtained by Fourier Transform Traction Cytometry (FTTC). (iii) Stress fields computed from FEM simulations using traction field as the external forces. (iv) Stress transformation gets the stress field oriented with the cell orientation.}
    \label{fig:fig1}
\end{figure}

\section{Methods}\label{sec:methods}

\subsection{Experimental methods}

\paragraph{Cell Culture}

Mouse myoblasts (C2C12; American Type Culture Collection, ATCC) and bone marrow--derived mesenchymal stem cells (BMSCs; Cell bank of CAS) were used in this study. All cells were cultured in Dulbecco’s Modified Eagle Medium--High Glucose (DMEM-HG; with L-glutamine and sodium pyruvate; Gibco), supplemented with 10\% fetal bovine serum (FBS; Gibco) and 1\% penicillin--streptomycin (100$\times$; Gibco). Cells were maintained in T75 culture flasks (Corning) under standard conditions (37~$^\circ$C, 5\% CO$_2$, humidified incubator), and the medium was replaced every 2--3 days. Subculturing was performed at $\sim$80\% confluency using 0.05\% trypsin--EDTA (Gibco). Cells were used between passages 10--20 to ensure stable proliferative and adhesive properties.

\paragraph{Substrate Preparation}

Substrates were prepared using polyacrylamide (PA) hydrogels with a stiffness of 21~kPa, controlled by adjusting the concentrations of acrylamide and bis-acrylamide during polymerization. To enable displacement tracking, carboxylate-modified fluorescent beads (200 nm, Invitrogen) were incorporated into the hydrogel solution before gelation. For cell adhesion, the hydrogel surfaces were functionalized with fibronectin via a protein transfer method: fibronectin was first adsorbed onto a glass coverslip (5~µL of 1~mg/mL fibronectin solution diluted into 200~µL H$_2$O), and PA gelation was carried out between this fibronectin-coated coverslip and an activated coverslip, forming a sandwich structure. After polymerization, the fibronectin-coated glass was carefully removed, leaving a uniform fibronectin layer on the hydrogel surface.

\paragraph{Traction force microscopy}

For traction force measurements on nematically ordered cells, cells were seeded onto polyacrylamide (PA) gel substrates at a density of $0.2 \times 10^6$ cells/well. Cells were allowed to attach, spread, and grow for several days until confluent monolayers with appropriate compactness formed. To achieve disordered states in C2C12 monolayers at approximately closed-cell density, cells were over-seeded at an initial density of $2 \times 10^6$ cells/well. In this case, cells were processed 4 hours after seeding to ensure complete attachment while preventing the onset of cellular rearrangement. Prior to TFM imaging, cell nuclei were live-labeled with Hoechst 33342 (Abcam). Fluorescent bead images were acquired using a Nikon Eclipse Ti-U fluorescence microscope equipped with $10\times$ objective lenses and a digital complementary metal-oxide semiconductor (CMOS) camera. Images were captured both before and after cell detachment induced by trypsinization to quantify substrate deformation.

\subsection{Computational framework }

We followed the workflow described by \cite{tambe_collective_2011TFM+MSM}, involving three main steps: (\textit{i}) acquisition of the substrate displacements $\Bu(\Br)$ ($\Br$ stands for the position vector), (\textit{ii}) computing the traction forces, $\BT(\Br)$, that cells exert on the soft elastic substrate, and (\textit{iii}) inferring the internal stresses within the cell monolayer by solving the elasticity problem. Next, we describe each of these processes in detail.

\paragraph{Analysis of Substrate Displacements}

%Since cell traction forces cause substrate deformations, cell detachment makes the substrate return to its reference (undeformed) state, resulting in a displacement field being measured. Following TFM, images of fluorescent beads acquired before and after cell detachment were compared via Particle Image Velocimetry (PIV) analysis, a cross-correlation–based technique for measuring flow fields. In PIV, consecutive image pairs are divided into interrogation windows, and the local displacement vector is determined by maximizing the cross-correlation between corresponding windows. We assumed small substrate displacements, such that linear elasticity applies.

Images of fluorescent microbeads embedded in the substrate were acquired before and after cell detachment with trypsin treatment to obtain the deformed and undeformed (reference) states, respectively. The displacement field was calculated using Particle Image Velocimetry (PIV) implemented in PIVLab in MATLAB. A three-pass cross-correlation scheme was applied. To remove rigid body motion, the mean values of the horizontal and vertical velocity components (\(u\) and \(v\)) were subtracted from the measured displacement field, followed by minimal smoothing.

\paragraph{Traction Force Reconstruction}

Given the measured substrate displacements $\Bu(\Br)$, the next goal is to infer the traction-stress field $\BT(\Br)$ from linear elasticity theory \cite{landau_elasticity}. Since the fluorescent bead displacements ($\sim5~\mu m$) are near the surface and much smaller than the substrate thickness ($\sim1~mm$), the substrate can be approximated as an isotropic elastic semi-infinite half-space. The core idea is to use the Green's function $\mathbf{G}(|\Br-\Br'|)$ and Fourier transform to solve for the traction field, which utilizes the relation between substrate displacement and traction stress expressed in Fourier space, and the method is known as Fourier Transform Traction Cytometry (FTTC) \cite{butler2002_fttc}. In this case, the Green's function is the Boussinesq solution to the linear differential equation for $\Bu(\Br)$, the displacement at point $\Br$, caused by $\BT(\Br')$, a point traction force at $\Br'$ on the surface of the elastic half-space. We have the displacement-traction relation:
\begin{equation}
    \Bu(\Br) = \mathbf{G}(|\Br-\Br'|)\BT(\Br').
\end{equation}
The superposition of effects from all traction forces gives the total displacements. Thus, the displacement field is written as a convolution of the Green's function (kernel) with the traction field,
\begin{equation}
    \Bu(\Br) = \mathbf{G}(|\Br-\Br'|)*\BT(\Br'),
\end{equation}
where $*$ denotes convolution. In 2D Fourier space, this becomes
\begin{equation}
    \tilde{\Bu}(\Bk) = \tilde{\mathbf{G}}(\Bk)\tilde{\BT}(\Bk)
\end{equation}
where the tilde denotes two-dimensional Fourier transform and $\Bk$ denotes wavevectors $(k_x, k_y)$ while $\tilde{\mathbf{G}}$ remains a $2\times2$ matrix. The solution to the traction field is thus given by matrix inversion in the Fourier space and inverse Fourier transform (denoted by $(\cdot)^{-FT}$) to the real space,
\begin{equation}
    \BT(\Br) = \left(\tilde{\mathbf{G}}^{-1}(\Bk)\tilde{\Bu}(\Bk)\right)^{-FT}.
\end{equation}

To ensure stability of the inversion, a Tikhonov regularization scheme was employed \cite{schwarz2002_regularization,kulkarni2018_regularization}, with the regularization parameter chosen to balance fidelity to the displacement data and suppression of noise. This procedure yields spatially resolved traction stresses at the cell–substrate interface. The parameters used in the ImageJ FTTC plugin were set as follows: Poisson’s ratio = 0.5, substrate stiffness = 21~kPa, and regularization factor = \(9 \times 10^{-9}\) \cite{tseng2012_fttc}.

\paragraph{Finite element simulation of internal stress}

To infer the internal stress distribution within the cell monolayer, we consider the cell confluent monolayer as a continuum elasticity problem. Assuming a state of mechanical equilibrium, substrate traction forces are balanced by internal stresses within the cellular sheet. We performed the simulation using the commercial software ABAQUS/2024. A 2D rectangular shell model is used with four-node bilinear plane-stress elements (CPS4R). The element size was chosen to be consistent with the grid size for traction force measurement (in this case $25~\mu m$). Traction forces obtained were imposed as surface traction forces (CLOAD) in the model. The cells are considered quasi-static at the moment of traction measurement, and thus we employed fixed boundary conditions for all four edges.

This integrated workflow thus provides a quantitative mapping from experimentally observed substrate displacements to the internal stress distribution across the cell monolayer. It allows us to bridge the scales between local cell–substrate interactions and emergent, tissue-scale stress organization.

\subsection{Data analysis}

\paragraph{Characterization of topological defect}

Topological defects in the two-dimensional orientation field were identified by calculating the winding number of director angles around a point. A closed contour $\Gamma$ was defined around the point--here a circular path of radius $R=100~\mu m$ was chosen--and the cell director angle $\theta$ was sampled at discrete points along the loop using bilinear interpolation. Because the nematic director is headless, the angles were first doubled and then unwrapped to ensure continuity before integration. The winding number is calculated as the net variation of angles along the contour $w=\sum_\Gamma \Delta\theta_i/(2\pi)$. A defect was identified with the value of winding number: $w=\frac{1}{2}$ and $w=-\frac{1}{2}$ correspond respectively to the characteristic comet-like and three-fold symmetric defect structures. Its spatial position was assigned to the geometric center of the sampling loop. Other magnitudes of the loop radius $R=50,200~\mu m$ were tested and obtained the same defect charge.

\paragraph{Quantification of nematic order}

The local nematic order parameter \(Q\) quantifies the degree of orientational alignment among nuclei or cellular directors within a defined region. 
For each subregion, the orientation angle of each element \(\theta_i\) was obtained by ellipse fitting or OrientationJ analysis, and the scalar order parameter was computed as
\[
Q = \sqrt{\langle \cos(2\theta_i) \rangle^2 + \langle \sin(2\theta_i) \rangle^2},
\]
where angular brackets denote averaging over all elements within the subregion. 
A value of \(Q=1\) corresponds to perfect alignment, while \(Q=0\) indicates a completely disordered orientation.

To assess spatial correlation of orientation, the two-point orientation correlation function \(C_{\theta\theta}(r)\) was computed as
\[
C_{\theta\theta}(r) = \langle \cos[2(\theta_i - \theta_j)] \rangle_{|r_i - r_j| = r},
\]
where \(\theta_i\) and \(\theta_j\) are the orientations of elements by distance \(r\). 
The averaging was performed over all pairs by distances within a narrow bin centered at \(r\). 
This correlation function characterizes the decay of orientational coherence with spatial separation.

\paragraph{Visualization of nematic order}
To quantify local orientational order of the cell monolayer, the discrete director field
$\theta(\mathbf{x}_i)$ obtained at each nodal point was converted into a complex nematic representation
$Z_i = e^{\,2\mathrm{i}\theta(\mathbf{x}_i)}$, where the double-angle formulation accounts for the head–tail symmetry of
nematic alignment. The real and imaginary parts of $Z_i$ were separately interpolated onto a regular spatial
grid using natural-neighbor interpolation to obtain a continuous complex field $Z(\mathbf{x})$. To suppress
local noise while preserving mesoscale features, $Z(\mathbf{x})$ was spatially smoothed by a normalized
Gaussian convolution,
\[
\tilde{Z}(\mathbf{x}) =
\frac{\displaystyle\int Z(\mathbf{x}')\,G_\sigma(\mathbf{x}-\mathbf{x}')\,d\mathbf{x}'}
     {\displaystyle\int G_\sigma(\mathbf{x}-\mathbf{x}')\,d\mathbf{x}'},
\]
where $G_\sigma$ is a Gaussian kernel of width $\sigma$. The local nematic order parameter was then computed
as the magnitude of the smoothed complex field,
\[
Q(\mathbf{x}) = \big|\tilde{Z}(\mathbf{x})\big|,
\]
yielding a continuous scalar map with $0 \le Q \le 1$, in which $Q=1$ denotes perfect local alignment and
$Q=0$ indicates isotropic disorder. 

\paragraph{Correlation analysis}
To quantify the relationship between factors (e.g. nucleus morphology, nematic order, nucleus orientation, director field, maximum principal stress), we performed correlation analyses using both Pearson and Spearman coefficients depending on the data type. Pearson’s \(r\) measures the strength of a linear relationship, while Spearman’s \(\rho_s\) evaluates monotonic association without assuming linearity. 
The Spearman coefficient is defined as
\[
\rho_s = 1 - \frac{6 \sum_i d_i^2}{N(N^2-1)},
\]
where \(d_i\) denotes the rank difference between paired observations and \(N\) is the number of samples. 
A higher value of \(\rho_s\) indicates a stronger positive correlation. The statistical significance was determined from the corresponding \(p\)-values.

\section{Results}

Understanding the interactions of contractile cells at cell-cell junctions is essential to explaining how local actomyosin contractility regulates the cells' collective behaviors such as alignment, migration, and morphogenesis. In confluent C2C12 myoblast monolayer, cells are elongated and highly polarized, often aligning into nematic-like patterns \cite{huo_myoblast_2025}. Such morphological anisotropy suggests that intercellular stresses are anisotropically distributed. To investigate this phenomemon, we reconstructed stress fields of C2C12 monolayers from TFM measurements following the procedures described in Sec.\ref{sec:methods} (Fig.~\ref{fig:fig1} and Fig.~\ref{figS:TFM}). This framework provides two major advantages. First, it allows mapping of stress distributions at subcellular resolution across the entire monolayer, making it possible to compare local stress anisotropy with cellular orientation. Second, it enables examination of collective stresses near structural irregularities such as topological defects, where nematic order breaks down. By quantifying both the magnitude and orientation of principal stresses, we can directly assess how tissue-level mechanical fields emerge from anisotropic cellular organization. This method not only reconstructs intra-monolayer stresses in a contractile cell sheets, but also establishes the foundation for evaluating how stress anisotropy aligns with nematic order and how stress reorganization contributes to morphological regulation in regions of disorder.

% Cells were cultured on compliant polyacrylamide substrates embedded with fluorescent particles, which served as fiducial markers for substrate deformation. The contractile activity of the monolayer displaced these markers, and the displacement field $u(x,y)$ was quantified from sequential images (Fig.~\ref{fig:fig1}c, left). Using Fourier transform traction cytometry (FFT-TC), the displacement field was converted into substrate traction stresses $T(x,y)$, representing the forces that cells exerted on the substrate (Fig.~\ref{fig:fig1}c, middle left).  

% To obtain the internal stress distribution within the monolayer, we solved the in-plane force balance equations by finite element method (FEM) using the traction stresses as boundary conditions. This yielded the stress tensor field $\sigma(x,y)$, including both normal stresses ($\sigma_{xx}, \sigma_{yy}$) and shear stresses ($\sigma_{xy}$) across the field of view (Fig.~\ref{fig:fig1}c, middle right). To analyze mechanical anisotropy, the stress tensor at each location was diagonalized to extract the principal stresses ($\sigma_{11}, \sigma_{22}$) and their orientation relative to the cell alignment (Fig.~\ref{fig:fig1}c, right).  

\begin{figure}[bt]
    \centering
    \includegraphics[width=0.9\linewidth]{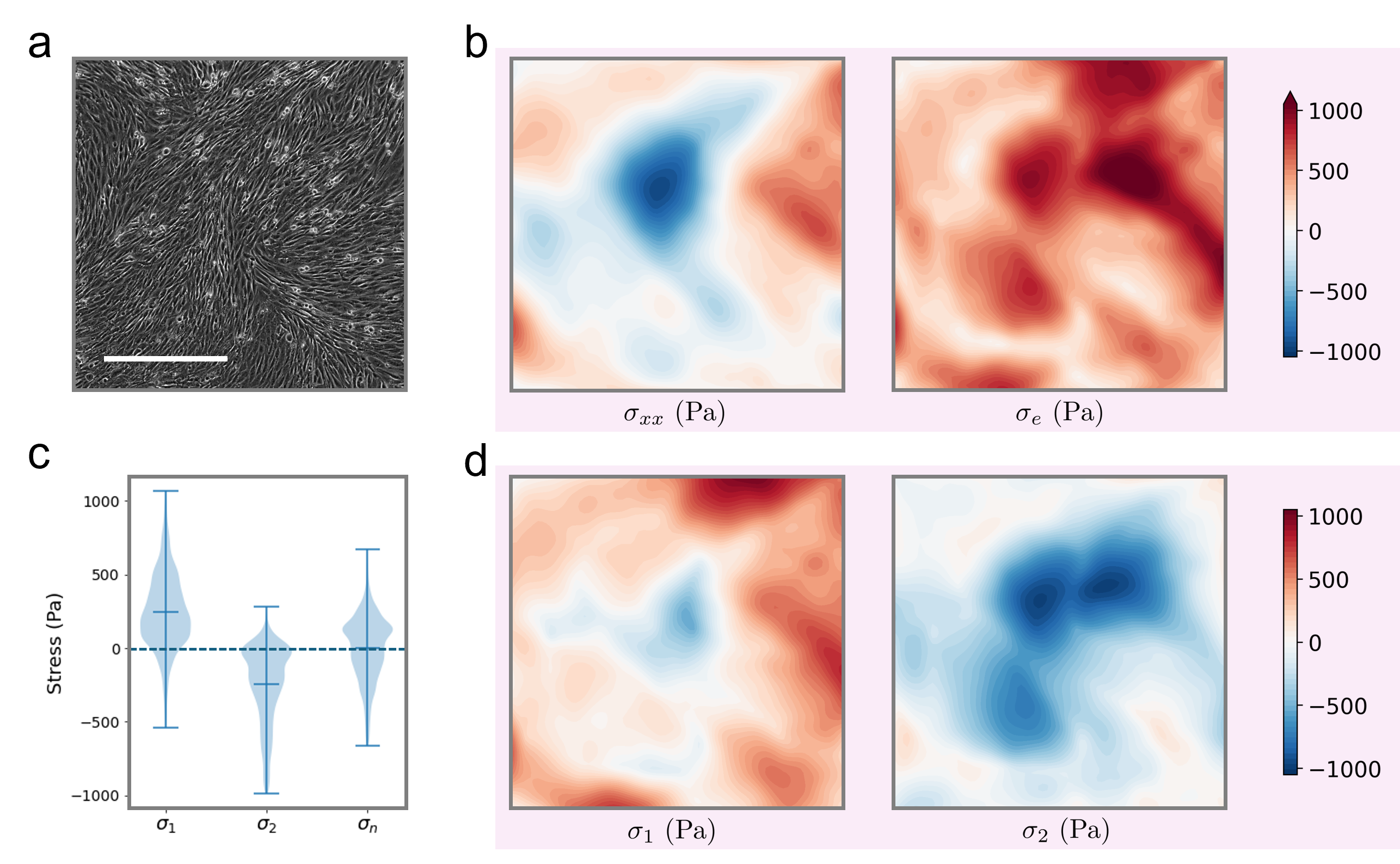}
    \caption{\textbf{Calculation of intra-monolayer stress using FEM.}
     \textbf{a}. Phase-contrast image of C2C12 cells. 
     \textbf{b}. Stress distribution for stress component in x-direction, $\sigma_{xx}$ (left), and effective stress, $\sigma_e$ (right), showing mechanical heterogeneity across the tissue. The color bar shows that red being tensile stress (positive), and blue being compressive stress (negative).
     \textbf{c}. Statistics of principal stresses ($\sigma_1$, $\sigma_2$) and mean normal stress ($\sigma_n$), displaying skewed distributions.
     \textbf{d}. Stress distribution for $\sigma_1$ and $\sigma_2$. Color bar indicates that $\sigma_1$ is dominantly tensile (positive), while $\sigma_2$ is primarily compressive (negative).
     Scale bar: 500~$\mu$m.}
    \label{fig:fem}
\end{figure}

\subsection{Stress distribution in C2C12 monolayer}

The confluent C2C12 monolayer is shown in Figure~\ref{fig:fem}a. Stess field was reconstructed  from the traction force data. The magnitude maps of the normal and effective stresses ($\sigma_{xx}$, $\sigma_{e}$) are shown in Figure~\ref{fig:fem}b (refer to Fig.~\ref{figS:other_stress_components} for other stress components). These maps reveal that the stress distribution across the monolayer is highly heterogeneous, reflecting local variations in contractility. We quantified the maximum and minimum principal stresses ($\sigma_{1}$, $\sigma_{2}$) as well as the mean normal stress ($\sigma_{n}$) in the monolayer (Fig.~\ref{fig:fem}c,d). Figure.~\ref{fig:fem}c shows that $\sigma_{1}$ was predominantly positive, with both the mean and median values being tensile across the monolayer,  consistent with the contractile nature of C2C12 cells. In contrast, $\sigma_{2}$ was largely negative (compressive). The anisotropic feature of the principal stresses suggests that the interactions between C2C12 cells strongly depended on their nematic orientation, consistent with their tendency to generate anisotropic tension, elongate, and alignment  \cite{huo_myoblast_2025}. It is noted that the mean normal stress, defined as $\sigma_n=\frac{1}{2}(\sigma_1+\sigma_2)$, exhibited an average value around zero (Fig.~\ref{fig:fem}c). This conclusion can be theoretically proved using the Divergence Theorem (see detailed explanation in SI section 1).  This stress distribution indicates that myoblasts generate anisotropic contractile forces, with tension dominating along one axis while compressive stresses arise along the other, reflecting a mechanically coordinated collective state.
%(\yc{Isn't this statment made according to Figure 4?}). 

\begin{figure}[tb]
    \centering
    \includegraphics[width=\linewidth]{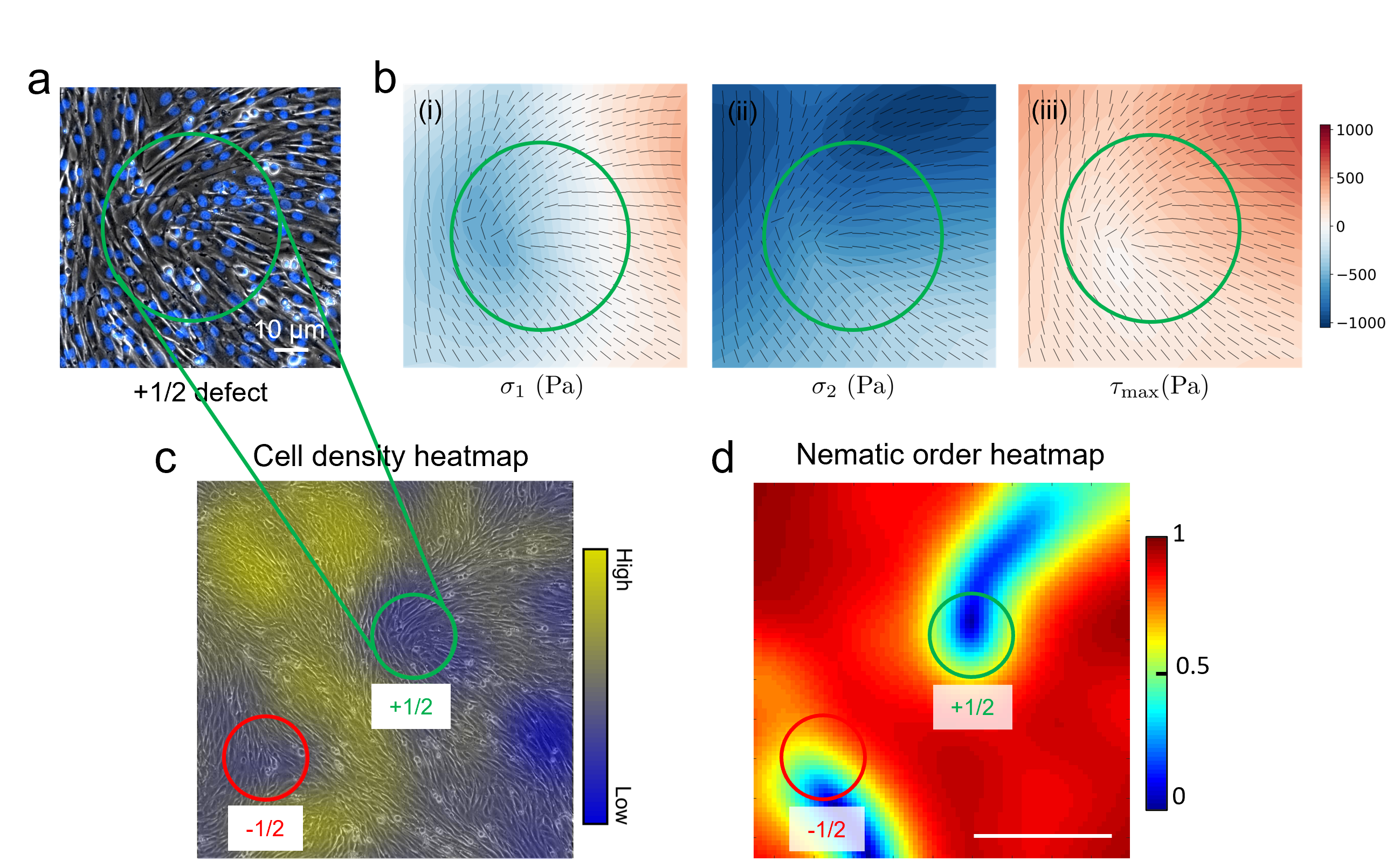}
    \caption{\textbf{Effects of +1/2 topological defect on the cellular and stress distribution.} 
    \textbf{a}. Zoom-in view of the defect with cell nucleus staining showing a local +1/2 topological defect. 
    \textbf{b}. Stress distributions near the +1/2 defect for derived principal stresses ($\sigma_1$ and $\sigma_2$) and maximum shear stress ($\tau_{\rm max}$). Bars indicate cell orientation directors. Green circles denote the same region in all figures.
    \textbf{c}. Cell density map overlaid with the monolayer phase-contrast image in Fig.~\ref{fig:fig1}a, with yellow color indicating higher density. The +1/2 topological defect is indicated in the green circle and corresponds to location of local low density.
    \textbf{d}. Heatmap of nematic order of the same field of view (FOV) in (c).
    Scale bar: 500~$\mu$m.}
    \label{fig:defect}
\end{figure}

\subsection{Topological defects as a mechanical and structural singularity}

Figure~\ref{fig:fem} shows the maximum principal stress exhibited an unusual presence of a local compressive region. Figure~\ref{fig:defect}a shows this region overlapped with a characteristic \(+1/2\) defect depicted by winding number analysis (Fig.~\ref{figS:Directors}). 
A magnified view of the area surrounding the defect (highlighted by the green circle in Fig.~\ref{fig:defect}a) reveals a distinct stress pattern: \(\sigma_1\) transitions from negative near the defect head to positive toward the tail. 
In contrast, \(\sigma_2\) remains compressive around the core while \(\tau_{\mathrm{max}}\) vanishes at the defect tip, indicating that the local stress state is nearly isotropic and the \(+1/2\) defect serves as a mechanical singularity where both stress and cell arrangement undergo a pronounced spatial transition (Fig.~\ref{fig:defect}b).

Figure~\ref{fig:defect}c shows that, when comparing the stress map with the corresponding cell density distribution, 
the topological defects consistently located in regions of reduced cell density, supporting that defects represent structural heterogeneity within the monolayer. 
The stress and density patterns identified here align with previous findings in active nematic systems, where \(+1/2\) defects self-propel from head to tail \cite{balasubramaniam_investigating_2021}. 
In our system, the tensile stress at the comet tail exerts a pulling effect on neighboring cells, while the compressive stress near the head drives local cell depletion.

We further examined the effects of topological defects on cell morphology by plotting spatial distribution of the nematic order parameter \(Q\), which quantifies local alignment of nucleus orientation (Fig.~\ref{fig:defect}d). 
It shows that the regions surrounding topological defects exhibit markedly lower \(Q\) values, indicating a disruption in nematic order. Notably, by analyzing a larger field of view (FOV) (Fig.~\ref{figS:AR_vs_Q}a), we demonstrated a strong positive correlation between local alignment (\(Q\)) and cell elongation (Fig.~\ref{figS:AR_vs_Q}b-d), as quantified by the mean nucleus aspect ratio (\(r_\mathrm{pearson}=0.78\), \(r_\mathrm{spearman}=0.80\)). 
Hence, cells in regions of low nematic order, where defects were located, were also less elongated. 
These results indicate that topological defects arise at locations where mechanical stress anisotropy diminishes, leading to reduced cell elongation and disrupted alignment, and revealing that local stress anisotropy serves as a mechanical cue that guides cell deformation and collective organization within the monolayer.

\subsection{Nematic organization governs stress anisotropy}

\begin{figure}[tbh]
\centering
\includegraphics[width=\linewidth]{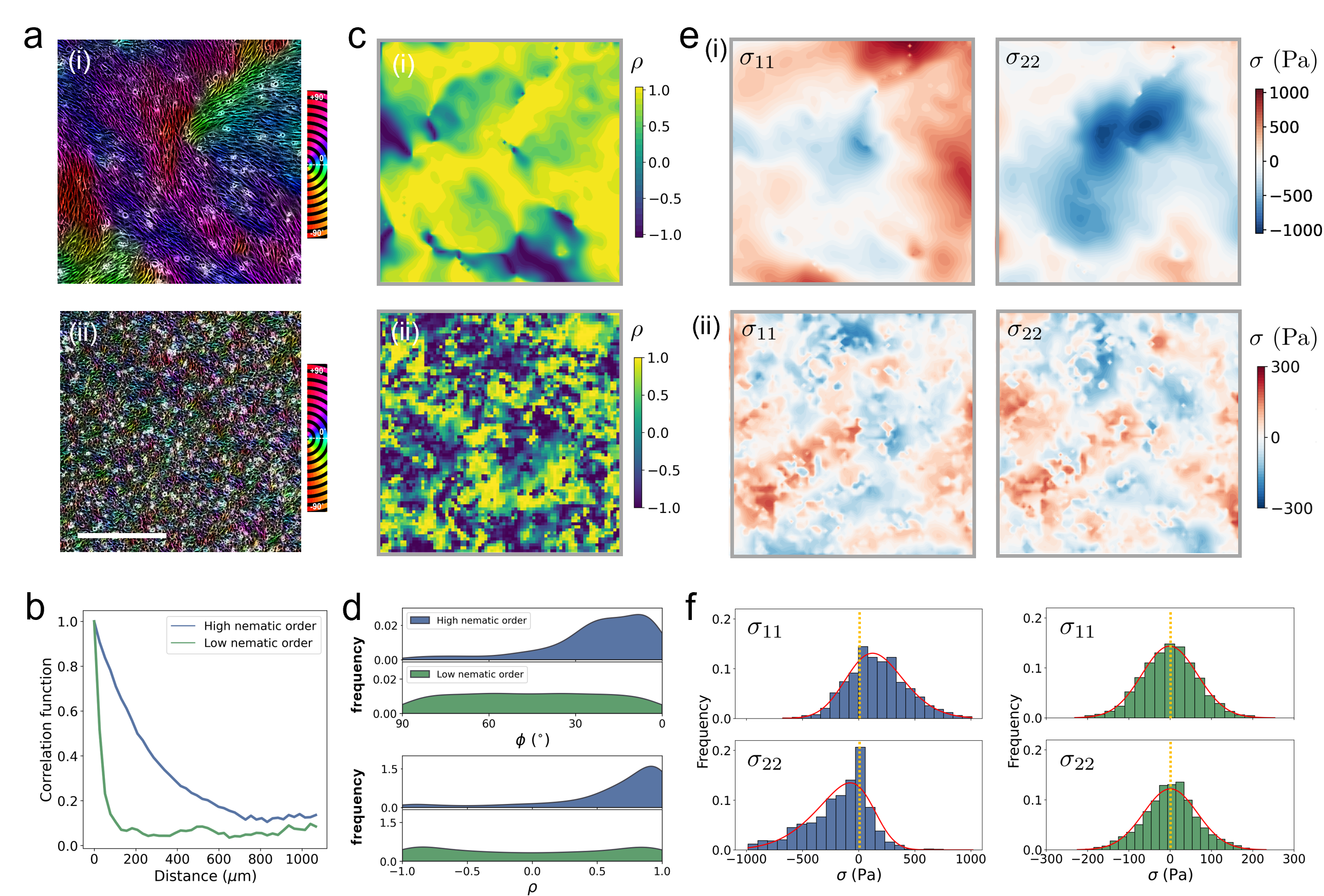}
    \caption{\textbf{Correlation between cell orientation and stress distribution.}
    \textbf{a}. Orientation maps of C2C12 monolayers with high nematic order (i) and low nematic order (ii). Colors indicate cell orientation angles ranging from $-90$ to $90^{\circ}$.
    \textbf{b}. Spatial correlation functions of the orientation fields for the two samples in \textbf{a}, highlighting the difference in correlation length. Blue and green curves correspond to the high and low nematic order cases, respectively.
    \textbf{c}. Spatial maps of correlation coefficient $\rho$ of the orientation field and maximum principal stress direction.
    \textbf{d}. Statistical distributions of the misalignment angle ($\phi$) and correlation coefficient ($\rho$) between the orientation field and the maximum principal stress direction in the two samples.
    \textbf{e}. Maps of stress components projected along the local cell orientation ($\sigma_{11}$) and perpendicular to it ($\sigma_{22}$).
    \textbf{f}. Statistical distributions of $\sigma_{11}$ and $\sigma_{22}$ for the two samples, showing distinct stress anisotropies.
    Scale bar: 500~$\mu$m.}
    \label{fig:4}
\end{figure}

Mechanical forces are tightly coupled with collective cell behaviors. In epithelial monolayers, traction forces correlate with velocity fields \cite{saraswathibhatla_EMLcoordinated_2021}, and cells tend to migrate along the direction of maximum principal stress \cite{tambe_collective_2011TFM+MSM}, a phenomenon known as plithotaxis \cite{trepat_plithotaxis_2011} where cells transmit normal stress across junctions while minimizing shear stress. In contrast, C2C12 myoblasts display weak cell--cell adhesion \cite{stehbens_perspectives_2024}, resulting in low velocity correlation ($C_{vv}$) \cite{huo_myoblast_2025,petitjean_velocity_2010}, yet maintain strong orientation correlation ($C_{\theta\theta}$) due to their elongated spindle shape and nematic order. This prompted the question whether intercellular stress also guides cell orientation in myoblast monolayers.  

To access this, we extracted the direction of maximum principal stress ($\theta_{\sigma_1}$) and compared it with the cell long-axis orientation ($\theta_{\text{cell}}$) by calculating the correlation coefficient

\[
\rho = \langle \cos 2\Delta \theta \rangle,
\]  

where $\Delta \theta = |\theta_{\sigma_1} - \theta_{\text{cell}}|$ and $\langle \cdot \rangle$ denotes averaging over all cells (demonstrated in Fig.~\ref{figS:cell_sigma1_alignment}). In nematically ordered monolayers, cells were highly elongated (Fig.~\ref{fig:4}a(i)), and $C_{\theta\theta}$ confirmed long-range alignment (Fig.~\ref{fig:4}b). Under these conditions, we found a strong alignment between $\sigma_1$ and the cell axis, with $\rho$ values close to 1 across most regions (Fig.~\ref{fig:4}c(i),d). This strong alignment between the direction of $\sigma_{1}$ and the cell long axis in ordered monolayers suggests that mechanical forces provide directional guidance for collective cell organization, analogous to the \textit{plithotaxis} behavior observed in epithelial tissues \cite{tambe_collective_2011TFM+MSM,trepat_plithotaxis_2011}. Here, rather than guiding migration, the stress field appears to orient cellular architecture: cells align their elongation axis with tensile stress directions. 
However, this guidance is highly sensitive to the cytoskeletal and contractile state of the monolayer. 
When cells were seeded at high density to suppress actin organization and intracellular contractility (Fig.~\ref{fig:4}a(ii) and Fig.~\ref{figS:ordered_vs_disordered}), they became less elongated (Fig.~\ref{figS:nucleus_comparison}a,c) and exhibited short-range $C_{\theta\theta}$ (Fig.~\ref{fig:4}b) and lower nematic order in nucleus arrangement (Fig.~\ref{figS:Q_comparison}). 
By reconstruction the stress inside disordered monolayer (Fig.~\ref{figS:stress_components_disordered}), we found this reduction in nematic order was accompanied by a weaker alignment between stress and cell axis, as reflected in the more scattered $\rho$ map (Fig.~\ref{fig:4}c(ii)) and lower average $\rho$ (Fig.~\ref{fig:4}d). 
These results highlight that the emergence of mechanical guidance requires an active, organized cytoskeletal network capable of responding to directional stresses.

To further elucidate how mechanical stresses align with cellular architecture, we decomposed the reconstructed stress tensor into the longitudinal ($\sigma_{11}$) and transverse ($\sigma_{22}$) components relative to the cell axis (shown in Fig.~\ref{fig:fig1}c). In nematically ordered monolayers, $\sigma_{11}$ was predominantly tensile whereas $\sigma_{22}$ remained largely compressive (Fig.~\ref{fig:4}e(i)), indicating that stresses are preferentially transmitted along the cell elongation axis. This anisotropic distribution of stress components reinforces the notion that cells both generate and align with tensile stresses, establishing a mechanically coherent network across the monolayer. In contrast, in disordered monolayers, this directional bias vanished: $\sigma_{11}$ and $\sigma_{22}$ exhibited nearly symmetric distributions around zero (Fig.~\ref{fig:4}e(ii), f), and other stress components displayed similar random patterns without spatial coherence (Fig.~\ref{figS:stress_components_disordered}).  

The distributions of $\sigma_{11}$ and $\sigma_{22}$ provide further insight into how the stress field couples to cell morphology (Fig.~\ref{fig:4}f). 
In ordered monolayers, $\sigma_{11}$ is biased toward tensile values while $\sigma_{22}$ is predominantly compressive, indicating that cells experience stretching along their elongation axis and compression transversely—consistent with their spindle-like shape. 
In contrast, in disordered monolayers, although the principal stresses ($\sigma_{1}$ and $\sigma_{2}$) remain positive and negative, respectively (Fig.~\ref{figS:stress_components_disordered}d-f), $\sigma_{11}$ and $\sigma_{22}$ exhibit nearly symmetric distributions around zero (Fig.~\ref{fig:4}f). 
This suggests that even though cells are still subjected to tensile force, the direction of the stress field is no longer aligned with the cell axis, consistent with the weak correlation between $\sigma_{1}$ and cell orientation observed for disordered cells (Fig.~\ref{fig:4}c,d). 
In disordered monolayers, tensile forces act transversely as well as longitudinally, effectively reducing cell elongation and promoting a more isotropic, circular morphology. 
This contrast between ordered and disordered states reveals how stress anisotropy governs cell deformation: when tensile stresses align with cell polarity, they reinforce elongation and collective order, whereas misaligned stresses disrupt anisotropy and drive morphological isotropization.

\subsection{Stress distribution for different contractile cell types}

\begin{figure}[h!]
\centering
\includegraphics[width=\textwidth]{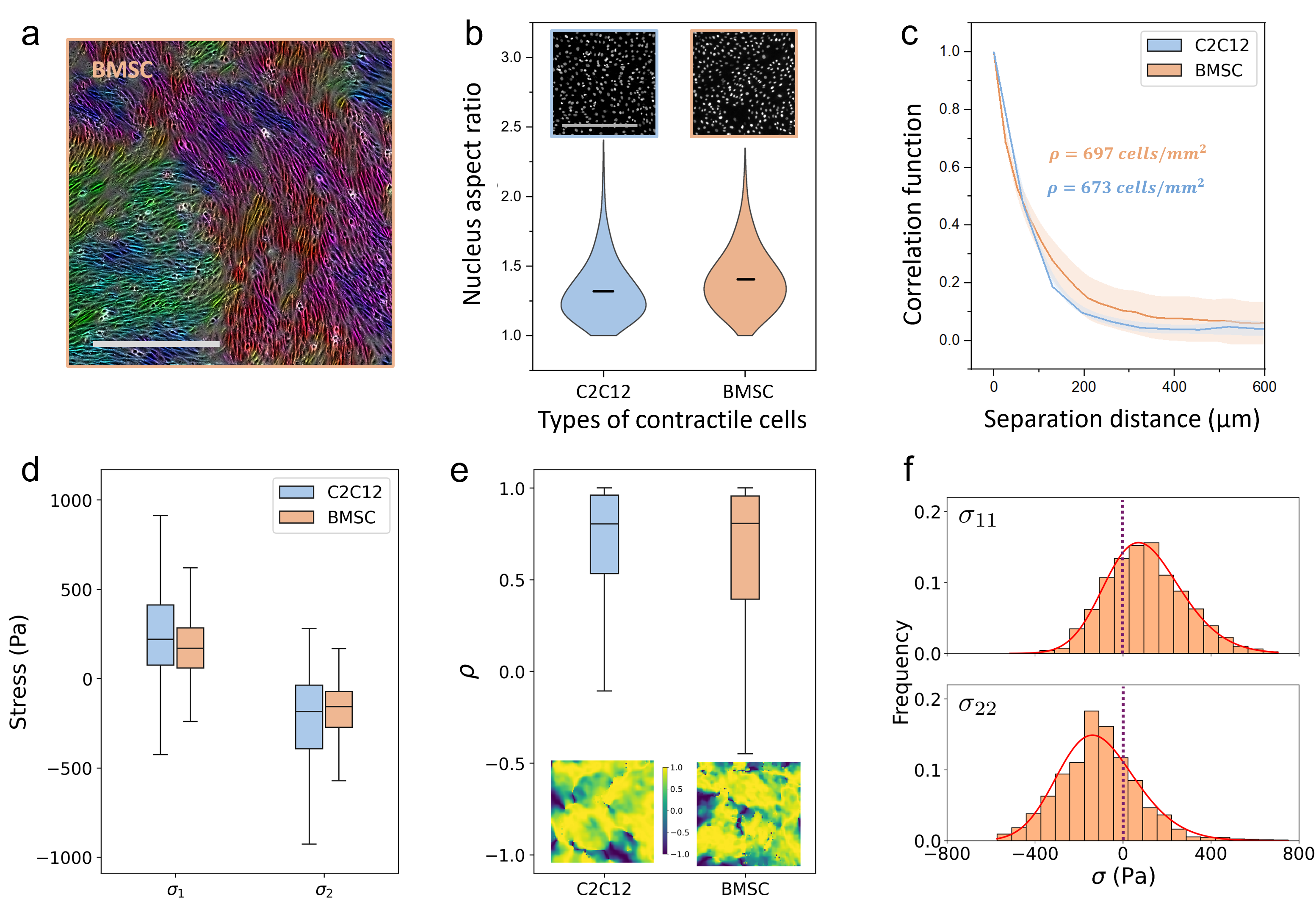}
\caption{
\textbf{Application of the stress analysis to another type of contractile cells.}
\textbf{a}. OrientationJ-derived orientation colormap of BMSCs. 
\textbf{b}. Comparison of nucleus aspect ratio between C2C12 myoblasts and BMSCs, showing that both exhibit elongated nuclear morphology. The inset shows the fluorescence images of nuclei. (n>1500 cells across 6 FOVs.)
\textbf{c}. Orientation correlation functions $C_{\theta\theta}(r)$ of C2C12 and BMSC monolayers. 
All comparisons were performed at comparable cell densities ($\rho_{\text{C2C12}} = 697~\mathrm{cells/mm^2}$, $\rho_{\text{BMSC}} = 673~\mathrm{cells/mm^2}$) to exclude the effects of spatial crowdedness. $C_{\theta\theta}(r)$ is averaged across 6 different FOVs.
\textbf{d}. Comparison of stress components $\sigma_1$ and $\sigma_2$ between C2C12 and BMSCs.
\textbf{e}. Distribution of correlation coefficient $\rho = \langle \cos 2(\theta_{\mathrm{cell}} - \theta_{\sigma_1}) \rangle$ between the cell orientation and the direction of $\sigma_{1}$, with representative spatial maps shown below. 
\textbf{f}. Histograms of longitudinal ($\sigma_{11}$) and transverse ($\sigma_{22}$) stress components of BMSC cells.
Scale bar: 500~$\mu$m.
}
\label{fig:5}
\end{figure}

In addition to C2C12 monolayers, we extended our analysis to other contractile-mesenchymal cell types, such as bone marrow--derived mesenchymal stem cells (BMSCs). 
BMSCs are multipotent progenitor cells capable of differentiating into various mesenchymal lineages, including osteoblasts, chondrocytes, and myocytes. 
They exhibit the hallmark features of mesenchymal and contractile phenotypes.
When cultured as confluent monolayers, BMSCs spontaneously developed orientationally coherent structures similar to those observed in C2C12 myoblasts (Fig.~\ref{fig:5}a). 
Despite differences in origin and differentiation potential, both cell types displayed an elongated nuclear morphology with comparable projected size (Fig.~\ref{figS:BMSC}d), and aspect ratios (Fig.~\ref{fig:5}b). Their orientation correlation functions $C_{\theta\theta}(r)$ revealed long-range nematic order with a similar correlation length (Fig.~\ref{fig:5}c). 
In the following, all comparisons between C2C12 and BMSC were conducted at matched cell densities to exclude the effects of cell packing or crowding.

Quantitative analyses revealed that, like C2C12 myoblasts, BMSC monolayer exhibits statistically positive $\sigma_{1}$ and negative $\sigma_{2}$, indicating its anisotropic stress state (Fig.~\ref{fig:5}d and Fig.~\ref{figS:BMSC}c). Meanwhile, it also shows a strong coupling between cell orientation and the direction of $\sigma_{1}$ (Fig.~\ref{fig:5}e) and a biased distribution of tensile $\sigma_{11}$ and compressive $\sigma_{22}$, reflecting the similar guidance of mechanical force on cell arrangement and morphology.
Together, these findings demonstrate that these mechanical characteristics are not unique to myoblasts but represents a universal mechanical principle of actively contractile, mesenchymal monolayers. Similarly, such high correlation between cell elongation and tensile stress direction would be disrupted in disordered systems, underscoring that  collective alignment and mechanical coherence are the outcome of intrinsic cytoskeletal contraction and cellualr polarity, rather than a cell-type–specific signaling or differentiation pathways.

\section*{Discussion}

An intriguing question arises from one of our most striking findings: how do local mechanical forces guide cellular arrangement? In ordered monolayers, we observed that cells tend to align with the direction of maximum principal stress (Fig.~\ref{figS:cell_sigma1_alignment} and Fig.~\ref{fig:4}c,d), revealing a clear mechanical guidance of collective organization. Yet, when cells lose their contractility and shape anisotropy, this guidance disappears—reflected in the disordered state, where stress and cell orientations become decoupled (Fig.~\ref{fig:4}c,d). This observation naturally raises an important question: does the director field obtained from OrientationJ truly represent the physical orientation of individual cells?

By comparing OrientationJ-derived directors with nucleus orientations from segmentation (Fig.~\ref{figS:cell_vs_director}a), we confirmed their strong correspondence in ordered monolayers (\(\rho = 0.7\); Fig.~\ref{figS:cell_vs_director}b) and weaker correlation in disordered ones (\(\rho = 0.19\)). 
Moreover, as shown in Fig.~\ref{figS:nucleus_director_AR}, this correspondence improved with increasing nucleus aspect ratio, indicating that elongated cells, by producing more anisotropic intensity patterns, are more faithfully captured by the structure-tensor algorithm of OrientationJ.  
When the true nucleus orientation was subsequently used to evaluate its coupling with the direction of \(\sigma_1\), the same trend persisted, i.e., ordered tissues exhibited strong alignment, whereas disordered ones did not (Fig.~\ref{figS:nucleus_sigma1}). 
This correlation also scaled with nucleus aspect ratio (Fig.~\ref{figS:nucleus_sigma1_AR}), underscoring that elongated cells are intrinsically more responsive to mechanical guidance. 
Together, these analyses confirm that the alignment between cell orientation and stress direction is not a computational artifact but a robust manifestation of force-guided organization—one that emerges only when cells possess sufficient elongation and contractile anisotropy to sense and respond to mechanical cues.

Interestingly, our finding in this work also resolves an open question from our previous study.
With the decomposition of the reconstructed stress tensor into components along and perpendicular to the cell axis, we found that, in nematically ordered monolayers, the longitudinal stress ($\sigma_{11}$) was predominantly tensile, whereas the transverse stress ($\sigma_{22}$) was compressive (Fig.~\ref{fig:4}e(i)), revealing a characteristic pattern of anisotropic tension.  
This quantitative result of anisotropic stress organization firmly supports our previous finding that the longitudinal adherens junctions are mechanically stronger and more persistent than the lateral ones \cite{huo_myoblast_2025}. This result further establishes a mechanistic connection between stress anisotropy and nematic order, i.e., the longitudinal forces transmit via focal adherens junction, maintain orientational coherence, and enable the coordinated contractility required for myogenic and other actively contractile cells.

We also examined the reliability of the reconstructed stress field, particularly regarding boundary conditions and field-of-view selection. 
Previous analyses \cite{tambe_MSMLimitation_2013} have demonstrated that, away from image edges, stresses reconstructed using fixed or symmetric boundaries are nearly identical, confirming that the boundary condition has minimal influence on interior stress distribution. 
With the fixed or symmetric boundary condition, our stress-reconstruction (Fig.~\ref{figS:BC_effect}) yields almost indistinguishable stress maps, while deviations appeared only when a single-edge fixed boundary condition is used. This validates the use of the fixed boundary condition, especially given that the cell displacement in our system is extremely small compared to the FOV size. 
Moreover, traction and stress reconstructions from different regions of the same sample (Fig.~\ref{figS:FOV}) exhibited consistent magnitudes and spatial patterns, confirming that our observations are intrinsic rather than dependent on boundary placement or region of interest selection. 
These validations collectively affirm the robustness of our analysis framework.

In summary, we developed and applied a computational framework combining traction force microscopy and finite element analysis to reconstruct the stress distributions in contractile cell monolayers. 
Our results reveal that stress anisotropy and nematic order are tightly coupled: stresses are transmitted preferentially along the cell elongation axis, and this alignment breaks down near topological defects where stress singularities emerge and cellular anisotropy diminishes. 
Furthermore, the observed correlation between stress anisotropy, cell shape, and nuclear morphology highlights the reciprocal interplay between mechanics and structure in collective cell organization. 
Together, these findings demonstrate that the self-organization of contractile monolayers arises from a fundamental physical principle—the mutual reinforcement of stress alignment and morphological order—that underlies the emergence of mechanical coherence in active biological tissues.

\section*{Authorship contribution statement}

Yucheng Huo: Conceptualization, Experimentats, Data curation, Formal analysis, Investigation, Methodology, Visualization, Writing - original draft, Writing – review \& editing. Kexin Guo: Conceptualization, Numerical simulation, Data curation, Formal analysis, Investigation, Methodology, Visualization, Writing - original draft, Writing – review \& editing. Massimo Paradiso: Formal analysis, Methodology, Writing. K. Jimmy Hsia: Conceptualization, Funding acquisition, Project administration, Supervision, Methodology, Writing - original draft, Writing – review \& editing, Funding acquisition.

\section*{Declaration of competing interest}

The authors declare no conflict of interest.

\section*{Acknowledgments}

YH, KG and KJH acknowledge the financial support of the Ministry of Education, Singapore, through the MOE AcRF Tier 3 Award (MOE-MOET32022-0002).

\section*{Data and code availability}

The data and code generated during the current study are available from the corresponding author upon request.

\section*{Declaration of use of generative AI and AI-assisted technologies}

During the preparation of this work, the authors used OpenAI’s ChatGPT-5 to improve the language and readability of the text. After using this tool, the authors reviewed and edited the content as needed and take full responsibility for its publication.

\bibliographystyle{elsarticle-num}
\bibliography{references}
\clearpage

% --- Start of Supplementary Information numbering ---
\setcounter{figure}{0}        % reset figure counter
\renewcommand{\thefigure}{S\arabic{figure}}  % prefix figures with "S"

\setcounter{table}{0}
\renewcommand{\thetable}{S\arabic{table}}    % (optional) prefix tables with "S"

\setcounter{equation}{0}
\renewcommand{\theequation}{S\arabic{equation}}  % (optional) prefix equations with "S"

% --- Start of Supplementary Information numbering ---
\setcounter{section}{0} % reset section numbering
\renewcommand{\thesection}{S\arabic{section}}

\setcounter{subsection}{0}
\renewcommand{\thesubsection}{S\arabic{section}.\arabic{subsection}}

% \documentclass[1p]{elsarticle}
% Use the lineno option to display guide line numbers if required.

% \usepackage{amsmath}
% \usepackage{amssymb}
% \usepackage{bm}
% \usepackage{graphicx}
% \usepackage[normalem]{ulem}

% %Macros to add comments
% \newcommand{\kg}[1]{\emph{\textcolor{magenta}{[#1]}}}%Macro for Kexin

% \newcommand{\Bepsilon}{{\boldsymbol{\varepsilon}}}
% \newcommand{\Bsigma}{{\boldsymbol{\sigma}}}
% \newcommand{\Bn}{{\mathbf{n}}}
% \newcommand{\Bu}{{\mathbf{u}}}
% \newcommand{\BI}{{\mathbf{I}}}
% \newcommand{\dA}{\,\mathrm{d}A}
% \newcommand{\ds}{\,\mathrm{d}s}
% \newcommand{\tra}{^{\sf T}}
% \DeclareMathOperator{\Tr}{tr}

\renewcommand{\theequation}{S\arabic{equation}}
\renewcommand{\thefigure}{S\arabic{figure}}
\renewcommand{\maketitle}{%
  \vskip45pt
  \begingroup
      \raggedright
      {\Large\bfseries Supporting Information for\par}
      \bigskip
      {\Large\bfseries Stress distribution in contractile cell monolayers \par}
      \bigskip
      {Yucheng Huo, Kexin Guo, Massimo Paradiso and K. Jimmy Hsia \par}
      \bigskip
      {Corresponding author: K. Jimmy Hsia \par E-mail: kjhsia@ntu.edu.sg}
  \endgroup
  \vskip45pt
  \section*{This PDF file includes:}
    \begin{list}{}{%
    \setlength\leftmargin{2em}%
    \setlength\itemsep{0pt}%
    \setlength\parsep{0pt}}
      \item Supporting text
      \item Figs.~S1 to S16
  \end{list}
  \clearpage
}

% \title{Supplementary Materials for Manuscript "Stress Distribution in Contractile Cell Monolayers"}

% \begin{document}

\maketitle

\newpage

\section{Derivation for average mean normal stress} 

In a plane stress condition the reduced Cauchy stress tensor $\Bsigma$ and the Green-Lagrange strain tensor $\Bepsilon$ are related by the 2D Hooke's law \cite{Timoshenko1969}:
\begin{equation}
	\Bsigma = \dfrac{E}{1-\nu^2}\big[\nu (\Tr{\Bepsilon}) \BI + (1-\nu) \Bepsilon \big] \,,
\end{equation}
where $E$ is the Young's modulus, $\nu$ is the Poisson's ratio, and $\BI$ is the 2D identity tensor. Hence, taking the trace of $\Bsigma$, the mean normal stress $\sigma_n$ results
\begin{equation}
	\sigma_n = \dfrac{\Tr{\Bsigma}}{2} = K_\mathrm{2D} \Tr{\Bepsilon} \,,
\end{equation}
with $K_\mathrm{2D} = E/(2(1-\nu$)) being the bulk modulus for the equivalent 2D material.

Recalling the linearization $\Bepsilon = (\nabla\Bu + (\nabla\Bu)\tra)/2$, and noting that $\Tr{\nabla\Bu} = \Tr{(\nabla\Bu)\tra} = \nabla\cdot\Bu$, the integral of $\sigma_n$ on the 2D domain $\Omega$ reads
\begin{equation}
	\int_{\Omega} \sigma_n \dA = K_\mathrm{2D} \int_{\Omega} \nabla\cdot\Bu \dA = K_\mathrm{2D} \int_{\partial\Omega} \Bu \cdot \Bn \ds \,,
\end{equation}
where the divergence theorem has been used, so that the fixed boundary condition implies the vanishing of the average normal stress:
\begin{equation}
	\dfrac{1}{A} \int_{\Omega} \sigma_n \dA = 0 \,.
\end{equation}
\clearpage

\section{Supplementary figures} 

\begin{figure}[h!]
\centering
\includegraphics[width=0.8\textwidth]{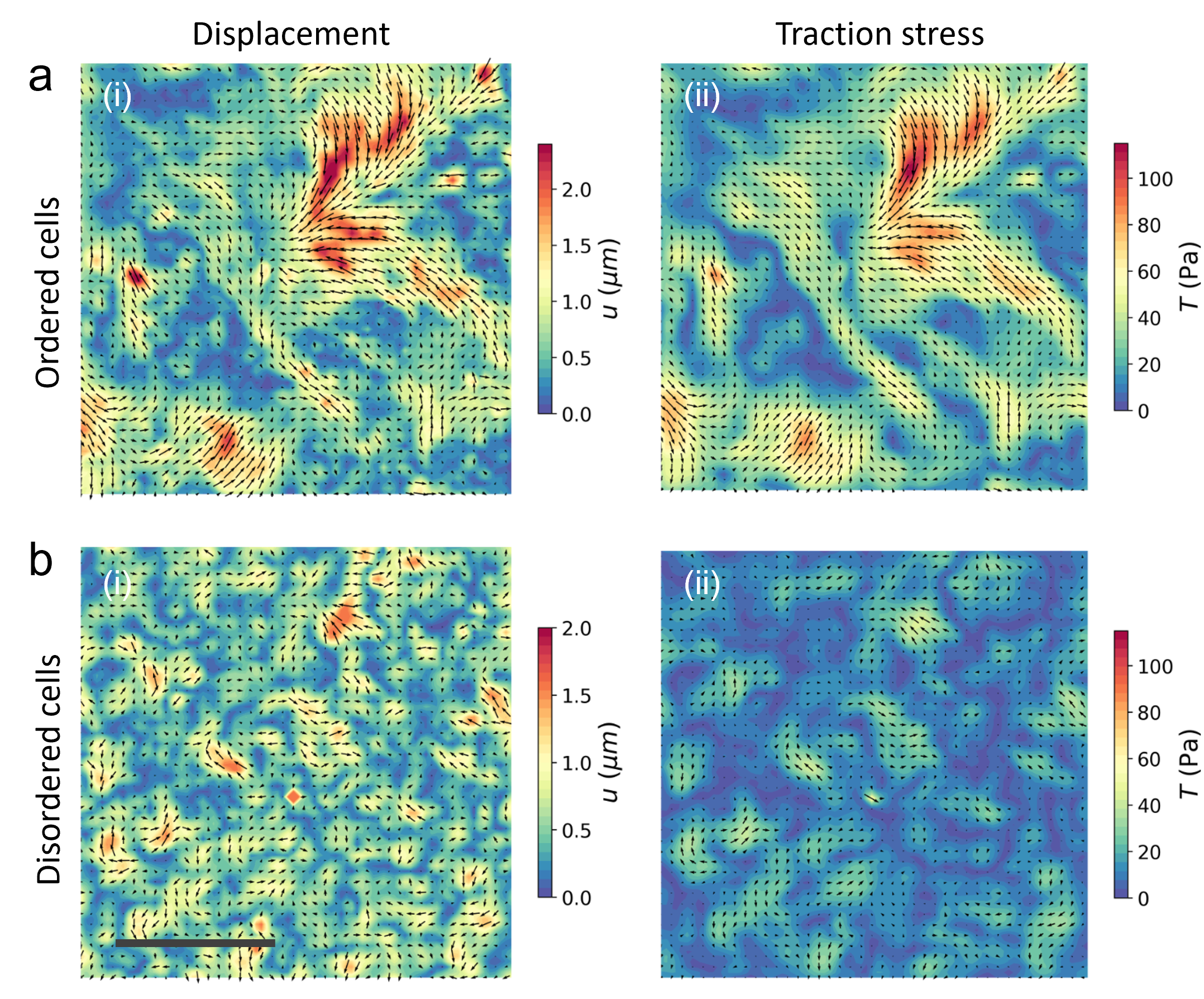}
\caption{
\textbf{TFM maps of nematically ordered and disordered C2C12 monolayers.} 
(i) and (ii) represent two independent fields of view for each condition. 
Color maps indicate traction stress magnitude, and black arrows denote traction vectors. 
Scale bar: 500~$\mu$m.
}
\label{figS:TFM}
\end{figure}
\clearpage

\begin{figure}[htbp]
\centering
\includegraphics[width=\textwidth]{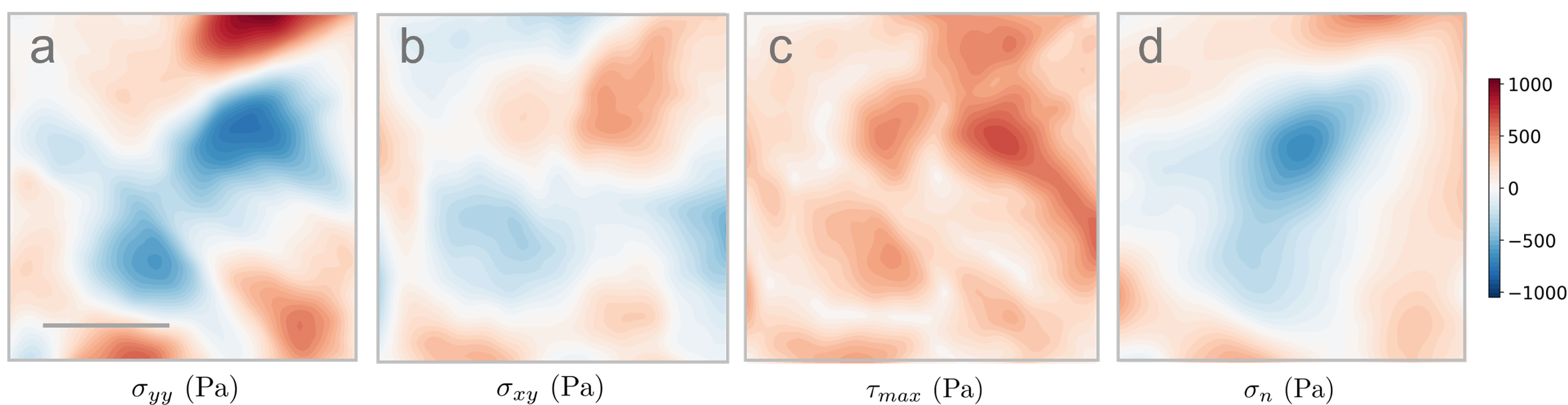}
\caption{\textbf{Additional stress components in ordered monolayers.}
(\textbf{a–d}) Spatial distributions of the remaining stress components obtained from the full stress tensor rotation corresponding to the ordered cell monolayer shown in Fig.~2. 
While Fig.~2 presents the primary components (\(\sigma_{xx}\), \(\sigma_{1}\), \(\sigma_{2}\), and \(\sigma_{e}\)), these maps illustrate the other rotated tensor components that together form the complete intra-monolayer stress field. 
Scale bar: \(500~\mu\mathrm{m}\).}
\label{figS:other_stress_components}
\end{figure}
\clearpage

\begin{figure}[htbp]
\centering
\includegraphics[width=0.7\textwidth]{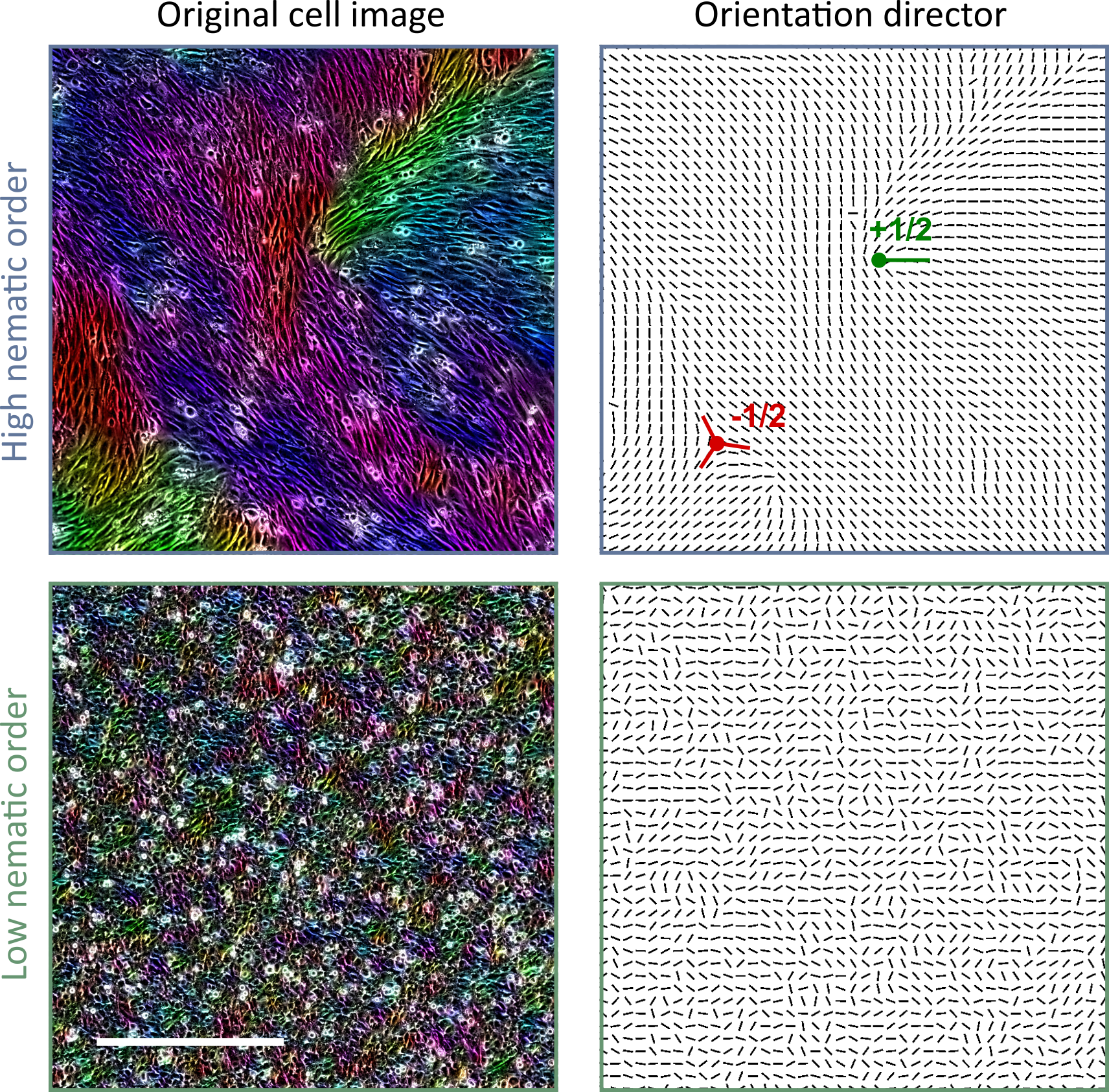}
\caption{\textbf{Characterization of cell oreintation.} 
Orientation colormap and director field obtained from OrientationJ analysis of ordered and disordered cells. Scale bar: \(500~\mu\mathrm{m}\).}
\label{figS:Directors}
\end{figure}
\clearpage

\begin{figure}[htbp]
\centering
\includegraphics[width=0.8\textwidth]{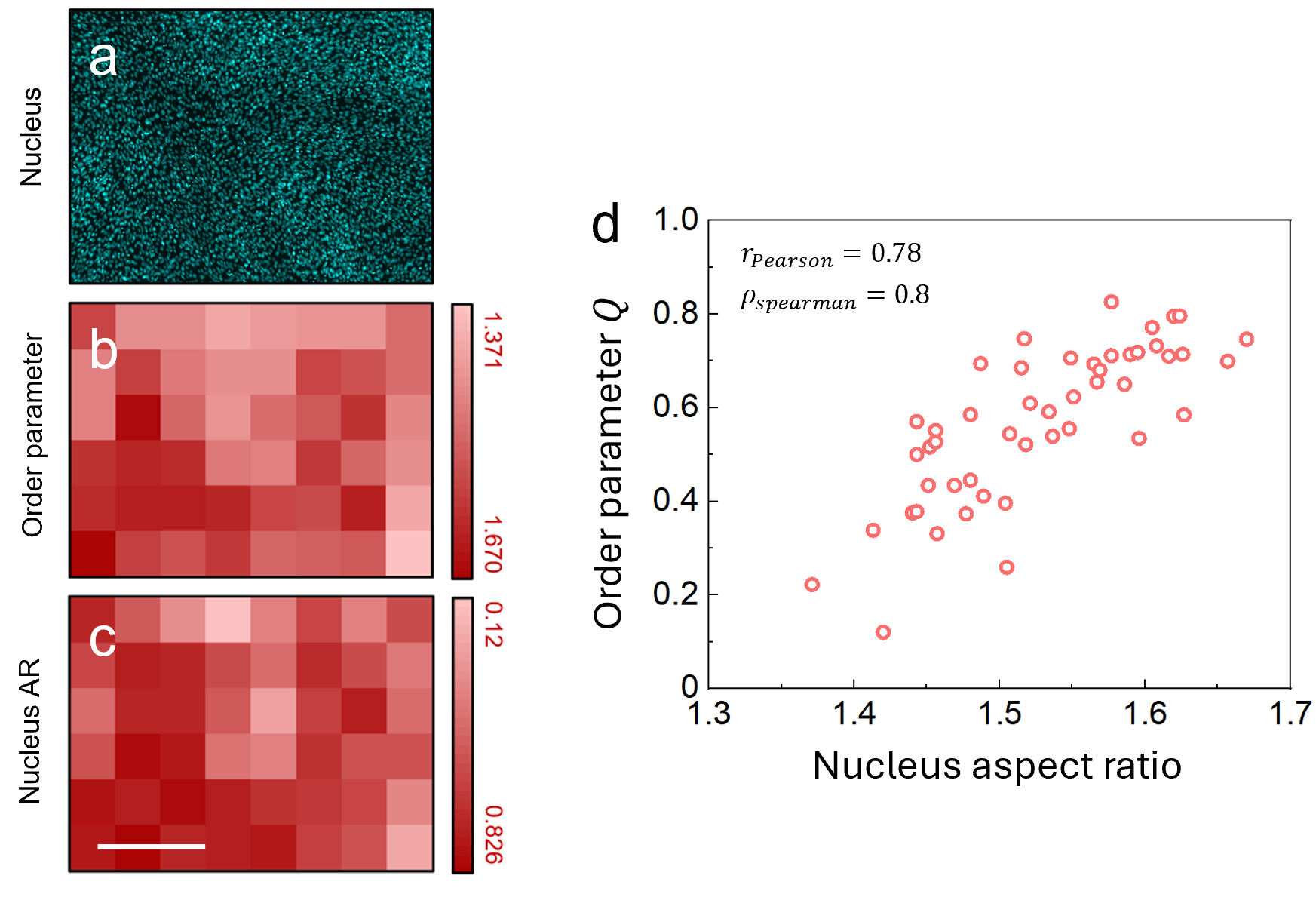}
\caption{\textbf{Correlation between nuclear aspect ratio and local nematic order.}
\textbf{a}. Representative fluorescence image of cell nuclei. 
\textbf{b}. Spatial map of the local nuclear aspect ratio (AR). 
\textbf{c}. Spatial map of the local order parameter (\(Q\)) computed from nuclear orientation within each subregion. 
\textbf{d}. Scatter plot showing the relationship between AR and \(Q\), revealing a strong positive correlation 
(Pearson’s \(r = 0.78\), \(p = 5.8\times10^{-11}\); Spearman’s \(\rho = 0.80\), \(p = 1.4\times10^{-11}\)). 
These results indicate that more elongated nuclei are associated with higher local alignment, confirming the coupling between cell shape anisotropy and nematic order. Scale bar: \(500~\mu\mathrm{m}\).}
\label{figS:AR_vs_Q}
\end{figure}
\clearpage

\begin{figure}[htbp]
\centering
\includegraphics[width=0.95\textwidth]{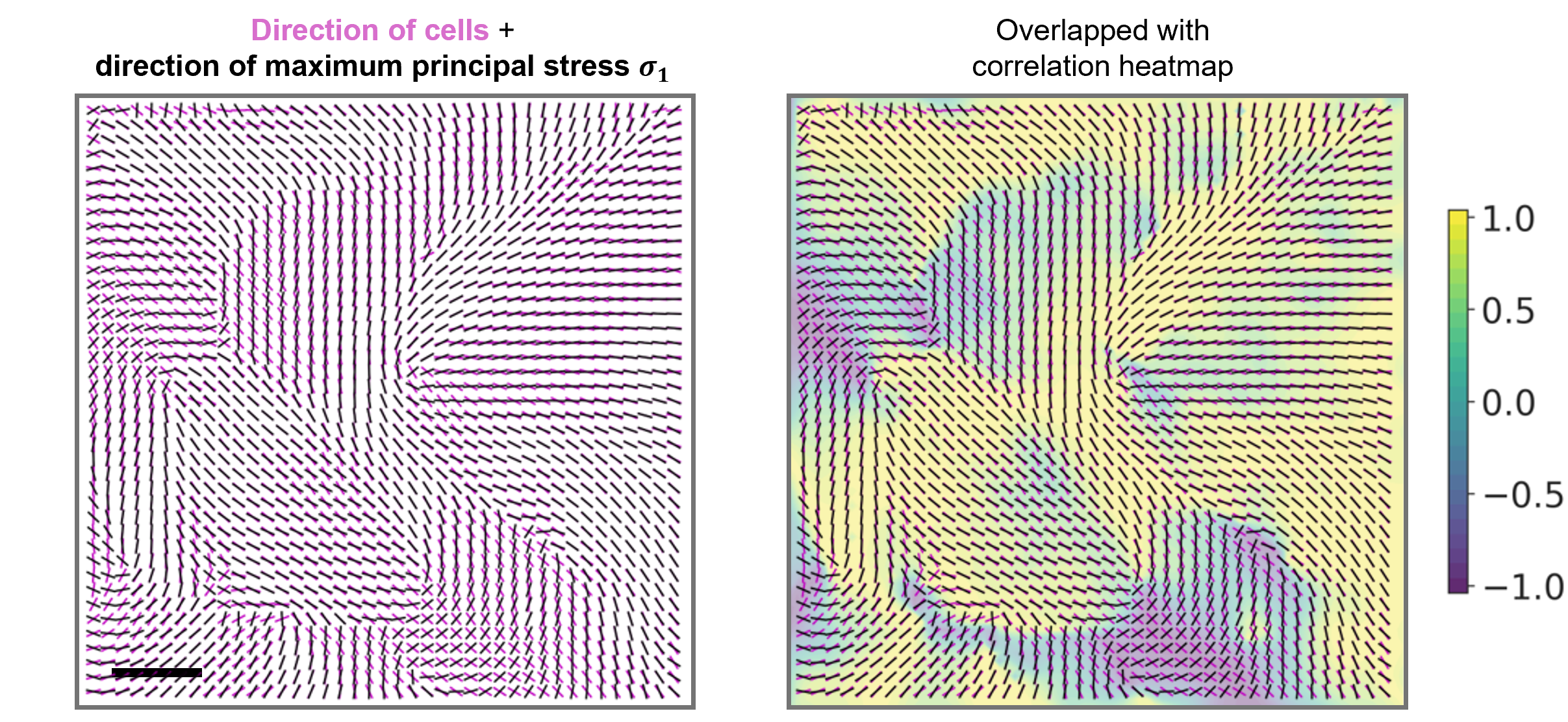}
\caption{\textbf{Visualization of cell orientation and principal stress direction in ordered monolayers.}
Director field of cells obtained from OrientationJ analysis (violet) on the left, and orientation field of the maximum principal stress \(\sigma_{1}\) (black). 
The director fields are overlap with heatmap of correlation between cell orientation and \(\sigma_{1}\) orientation \(\rho_{\mathrm{cell}-\sigma_{1}}\) shown in Figure~4c. The regions where directors of cell and \(\sigma_{1}\) prependicular align well with the dark regions in correlation coefficient heatmap, validating the computational method. 
Scale bar: \(200~\mu\mathrm{m}\).}
\label{figS:cell_sigma1_alignment}
\end{figure}
\clearpage

\begin{figure}[htbp]
    \centering
    \includegraphics[width=\textwidth]{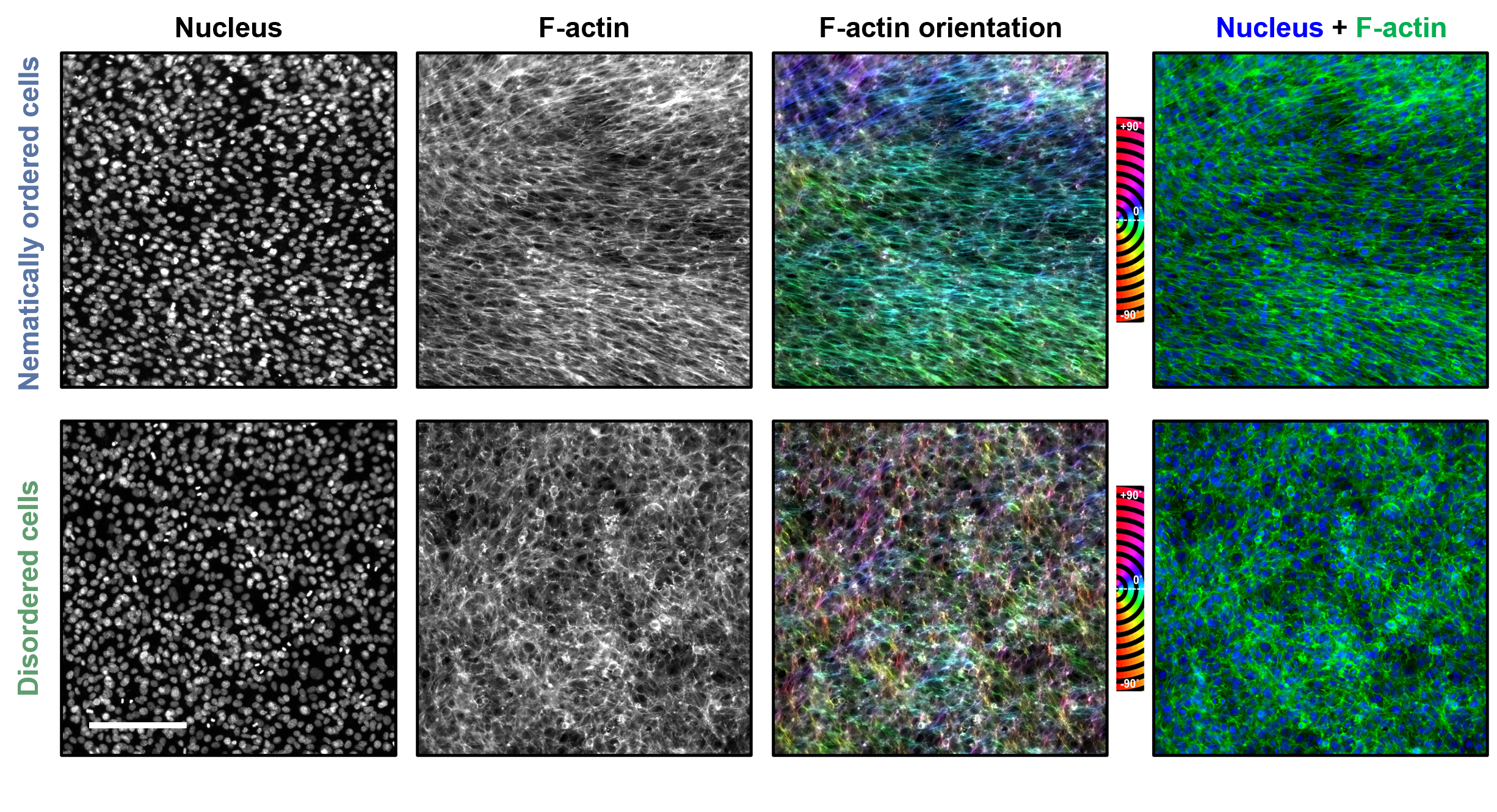}
    \caption{
    \textbf{Representative fluorescence images comparing nematically ordered and disordered C2C12 monolayers.} 
    Ordered cells (top row) exhibit elongated nuclei, aligned actin fibers, and coherent orientation domains, indicating strong cytoskeletal organization and contractility. 
    In contrast, disordered cells (bottom row) display rounder nuclei, isotropic actin architecture, and loss of alignment, consistent with reduced contractility and diminished nematic order. 
    Nuclei and F-actin are shown in blue and green. Scale bar: 200~$\mu$m.
    }
    \label{figS:ordered_vs_disordered}
\end{figure}
\clearpage

\iffalse
\begin{figure}[htbp]
\centering
\includegraphics[width=\textwidth]{Figures/FigS_sigma11_and_22_of_ordered_and_disordered_cells.png}
\caption{\textbf{Distributions of normal stress components in ordered and disordered monolayers.}
Histograms of longitudinal (\(\sigma_{11}\)) and transverse (\(\sigma_{22}\)) stress components for nematically ordered (left) and disordered (right) cells. 
Ordered tissues exhibit higher tensile \(\sigma_{11}\) and more compressive \(\sigma_{22}\), indicating pronounced anisotropic tension, whereas disordered tissues show more symmetric and isotropic stress distributions.}
\label{figS:sigma11_sigma22_distribution}
\end{figure}
\clearpage
\fi

\begin{figure}[htbp]
\centering
\includegraphics[width=\textwidth]{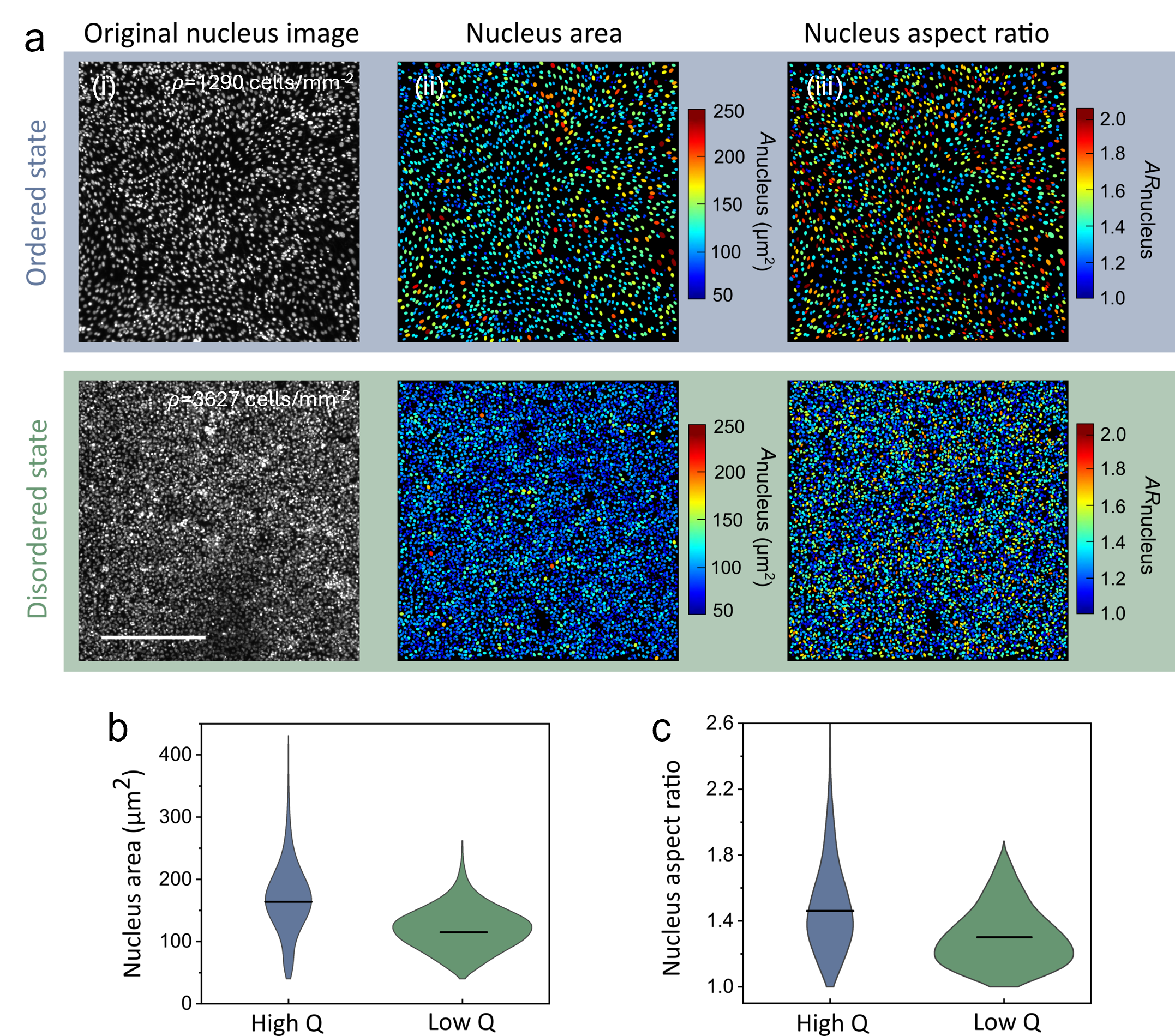}
\caption{\textbf{Comparison of nucleus morphology between the analyzed FOVs of ordered and disordered monolayers.}
\textbf{a}. (\textit{i}) Representative fluorescence images of nuclei in ordered and disordered states, showing distinct cell densities (\(\rho = 1290\) and \(3627~\mathrm{cells/mm^2}\), respectively). (\textit{ii, iii}) Corresponding color maps of nucleus area (\(A_{\mathrm{nucleus}}\)) and aspect ratio (\(AR_{\mathrm{nucleus}}\)) for the same FOV. Ordered monolayers, characterized by nematic alignment, display larger and more elongated nuclei, consistent with lower cell density and anisotropic cell stretching. In contrast, disordered monolayers exhibit smaller, more isotropic nuclei. 
\textbf{b, c}. The violin plots quantify these differences, showing significantly larger \(A_{\mathrm{nucleus}}\) and higher \(AR_{\mathrm{nucleus}}\) in the ordered state compared to the disordered state. Scale bar: \(500~\mu\mathrm{m}\).}
\label{figS:nucleus_comparison}
\end{figure}
\clearpage

\begin{figure}[htbp]
\centering
\includegraphics[width=0.6\textwidth]{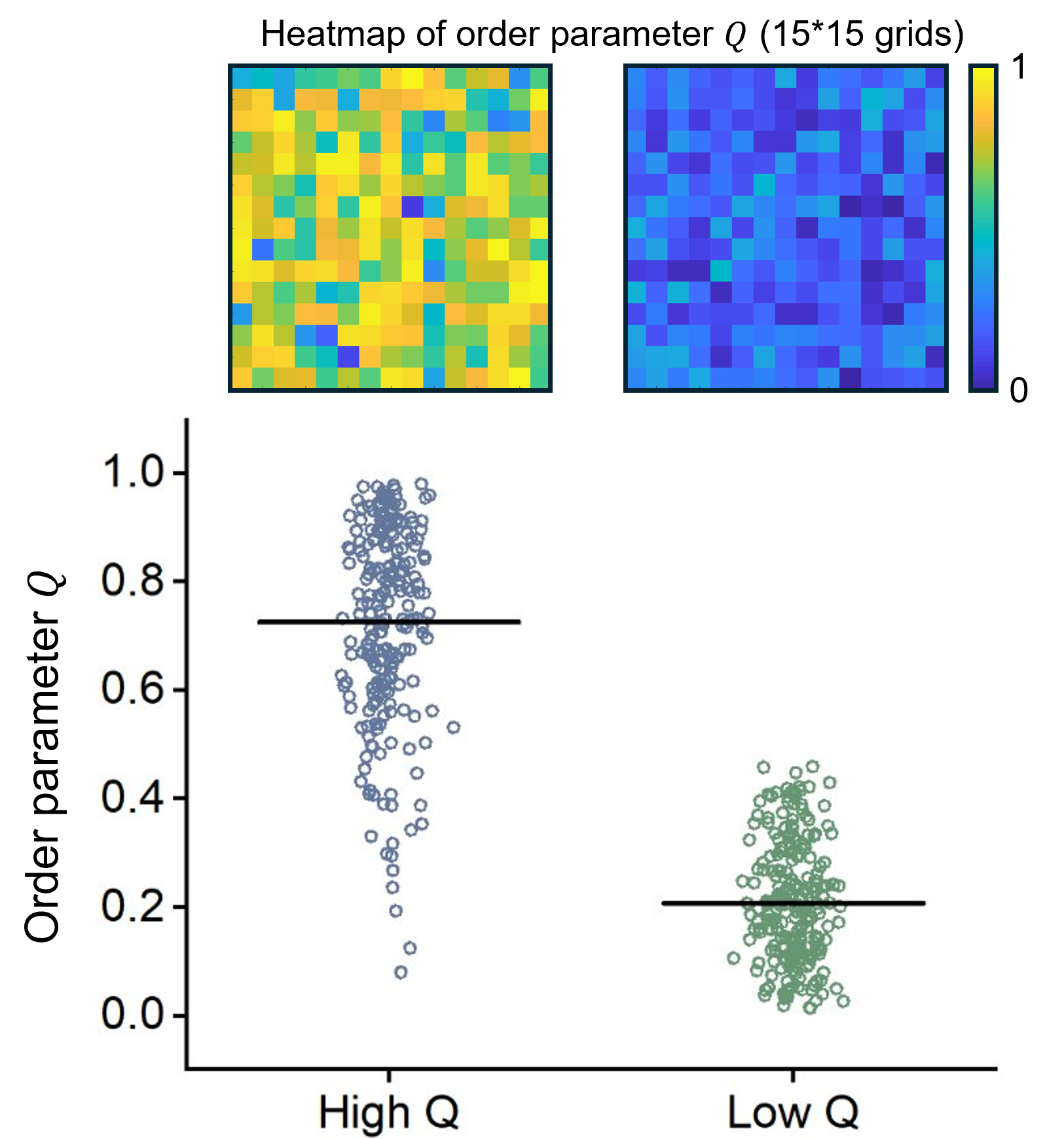}
\caption{\textbf{Quantification of nematic order in ordered and disordered cell monolayers.}
Representative heatmaps of the spatial distribution of the local nematic order parameter \( Q \) computed from nucleus orientation in (left) ordered and (right) disordered cell monolayers. Each pixel corresponds to the mean order parameter within a subregion of the FOV. The plot below compares the distributions of \( Q \) values between the two states, showing significantly higher \( Q \) in the ordered (high-\(Q\)) monolayer compared to the disordered (low-\(Q\)) monolayer, confirming distinct collective alignment behaviors.}
\label{figS:Q_comparison}
\end{figure}
\clearpage

\begin{figure}[htbp]
\centering
\includegraphics[width=\textwidth]{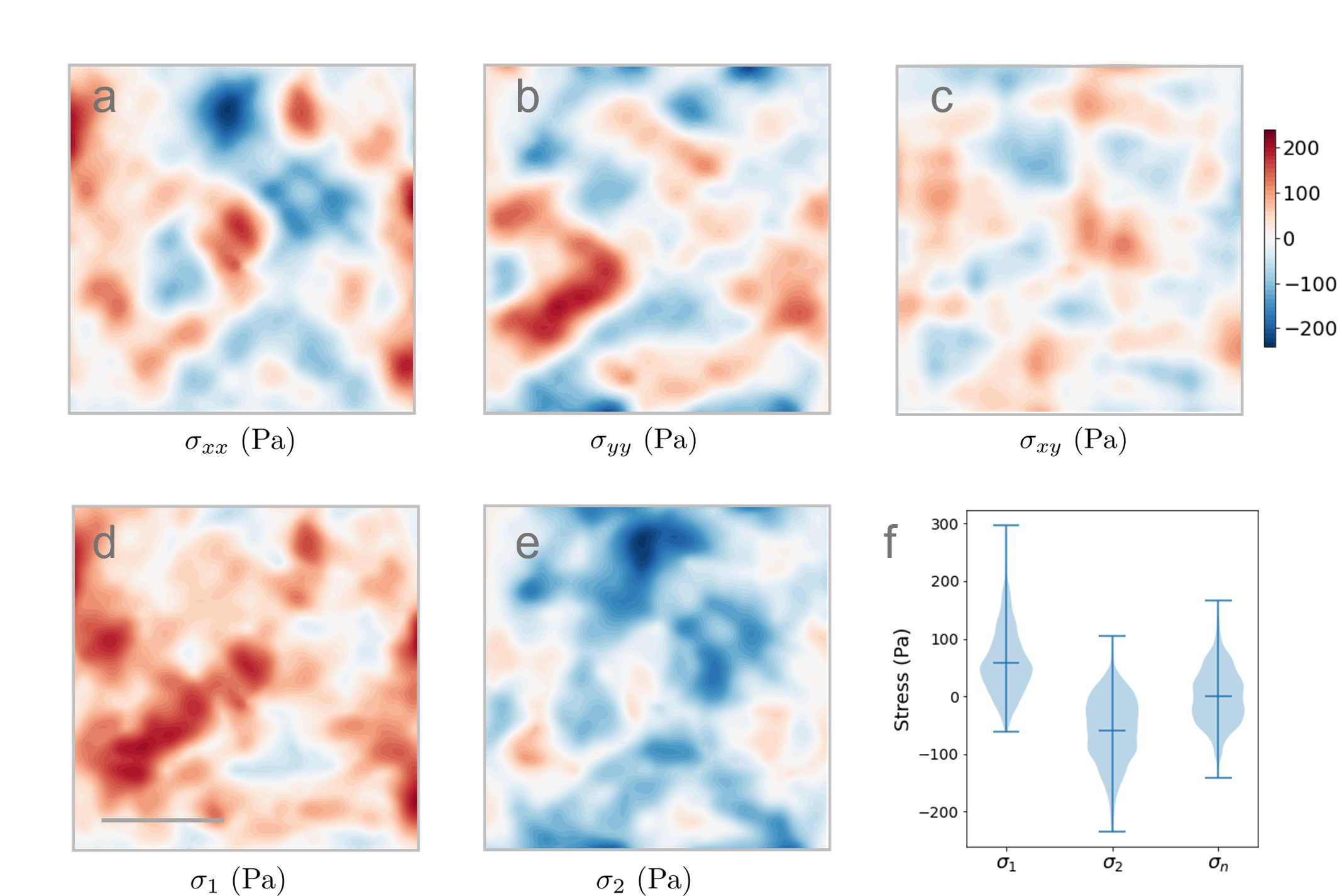}
\caption{\textbf{Stress components in disordered monolayers.}
\textbf{a–e}. Spatial maps of the stress components reconstructed from the full stress tensor of the disordered cell monolayer presented in Fig.~4. 
While Fig.~4 shows the main normal stresses (\(\sigma_{11}\) and \(\sigma_{22}\)), here we display the remaining tensor components that together describe the complete intra-monolayer stress field. 
Specifically, \(\sigma_{xx}\) and \(\sigma_{yy}\) denote the normal stresses along the laboratory \(x\)- and \(y\)-axes, respectively; 
\(\sigma_{xy}\) and \(\sigma_{yx}\) represent the shear stresses that characterize local tangential deformations; 
\(\sigma_{1}\) and \(\sigma_{2}\) correspond to the maximum and minimum principal stresses, respectively; 
and \(\sigma_{e}\) indicates the effective (von Mises–type) stress magnitude. 
\textbf{f}. Violin plots summarize the statistical distribution of stress magnitudes for the different components: \(\sigma_{1}\), \(\sigma_{2}\), and mean normal stress $\sigma_n$. 
Scale bar: \(500~\mu\mathrm{m}\).}
\label{figS:stress_components_disordered}
\end{figure}
\clearpage

\begin{figure}[h!]
\centering
\includegraphics[width=0.8\textwidth]{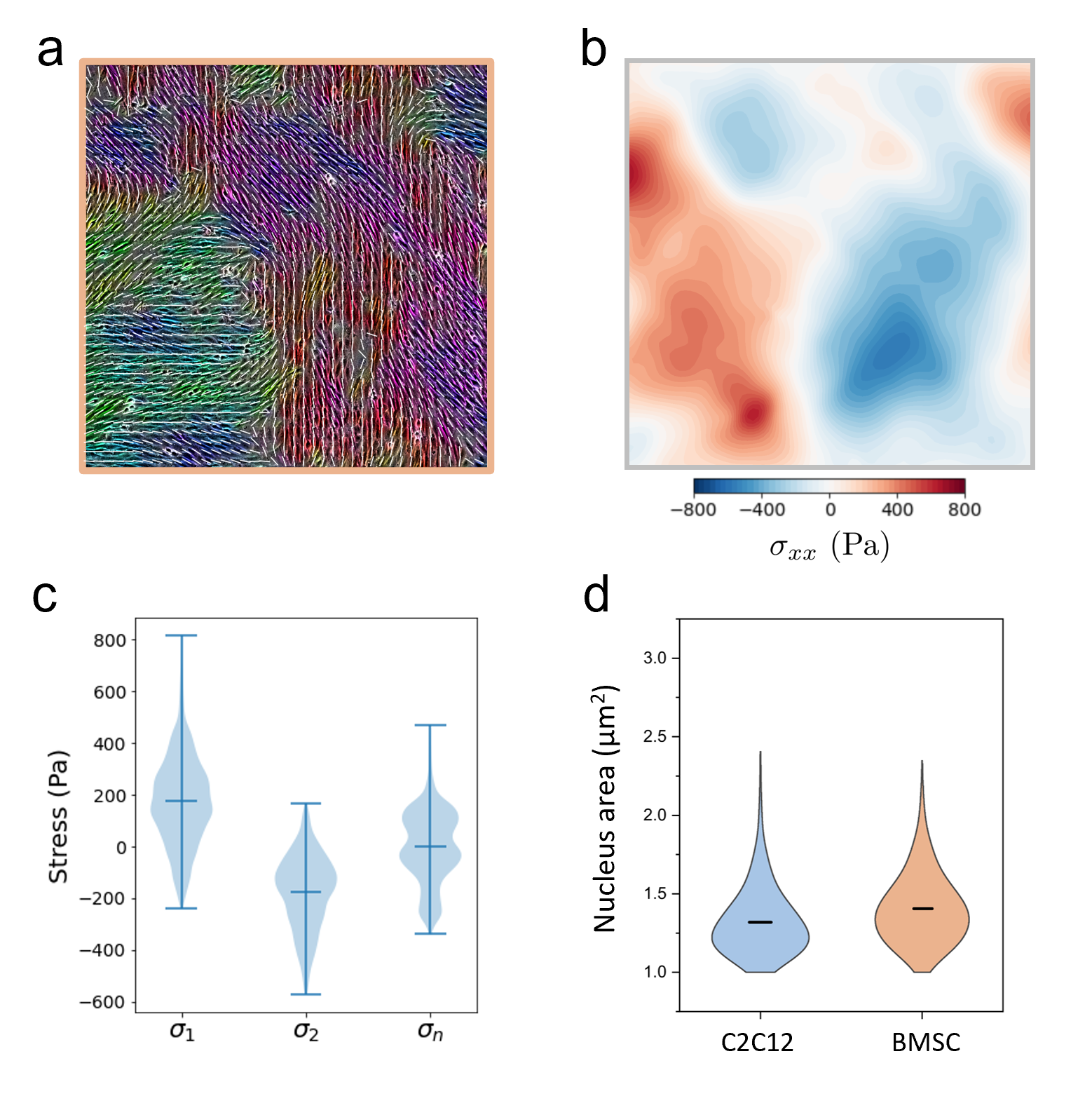}
\caption{
\textbf{Characterization of mechanical anisotropy in bone marrow--derived mesenchymal stem cell (BMSC) monolayers.} 
\textbf{a}.  Spontaneous nematic order of BMSCs, visualized by the OrientationJ–derived director field overlaid on the orientation colormap. 
\textbf{b}. Representative reconstructed stress map ($\sigma_{xx}$ component) of BMSC monolayer obtained from MSM. 
\textbf{c}. Distribution of stress components \(\sigma_{1}\), \(\sigma_{2}\), and \(\sigma_{n}\) of BMSCs.
\textbf{d}. Comparison of nucleus projected area between C2C12 and BMSC monolayers (n>1500 cells).
}
\label{figS:BMSC}
\end{figure}
\clearpage

% \iffalse
% \begin{figure}[htbp]
% \centering
% \includegraphics[width=0.8\textwidth]{Figures/FigS_autocorrelation_heatmap.png}
% \caption{\textbf{Autocorrelation of the longitudinal normal stress \(\sigma_{11}\) in ordered and disordered monolayers.}
% (\textbf{a}) Ordered monolayer with high nematic order shows an anisotropic, diamond-shaped correlation pattern, indicating coherent stress propagation along the nematic axis. 
% (\textbf{b}) Disordered monolayer exhibits an isotropic and rapidly decaying correlation, reflecting the loss of long-range mechanical coherence.}
% \label{figS:autocorr_sigma11}
% \end{figure}
% \clearpage
% \fi

% \iffalse
% \begin{figure}[htbp]
% \centering
% \includegraphics[width=0.8\textwidth]{Figures/FigS_autocorrelation_function_curve.png}
% \caption{\textbf{Radially averaged spatial correlation of intra-monolayer stress.}
% (\textbf{a}) Comparison of the normalized autocorrelation function \(C(r)\) of the longitudinal stress \(\sigma_{11}\) between ordered and disordered monolayers. 
% The ordered state exhibits a slower decay of \(C(r)\), indicating longer-range stress coherence. 
% (\textbf{b}) Radial correlation functions of different stress components in the ordered state, showing similar decay behavior among the principal and normal stresses (\(\sigma_1\), \(\sigma_{2}\), \(\sigma_{11}\), \(\sigma_{22}\)).}
% \label{figS:autocorr_curve}
% \end{figure}
% \clearpage
% \fi

\begin{figure}[htbp]
\centering
\includegraphics[width=\textwidth]{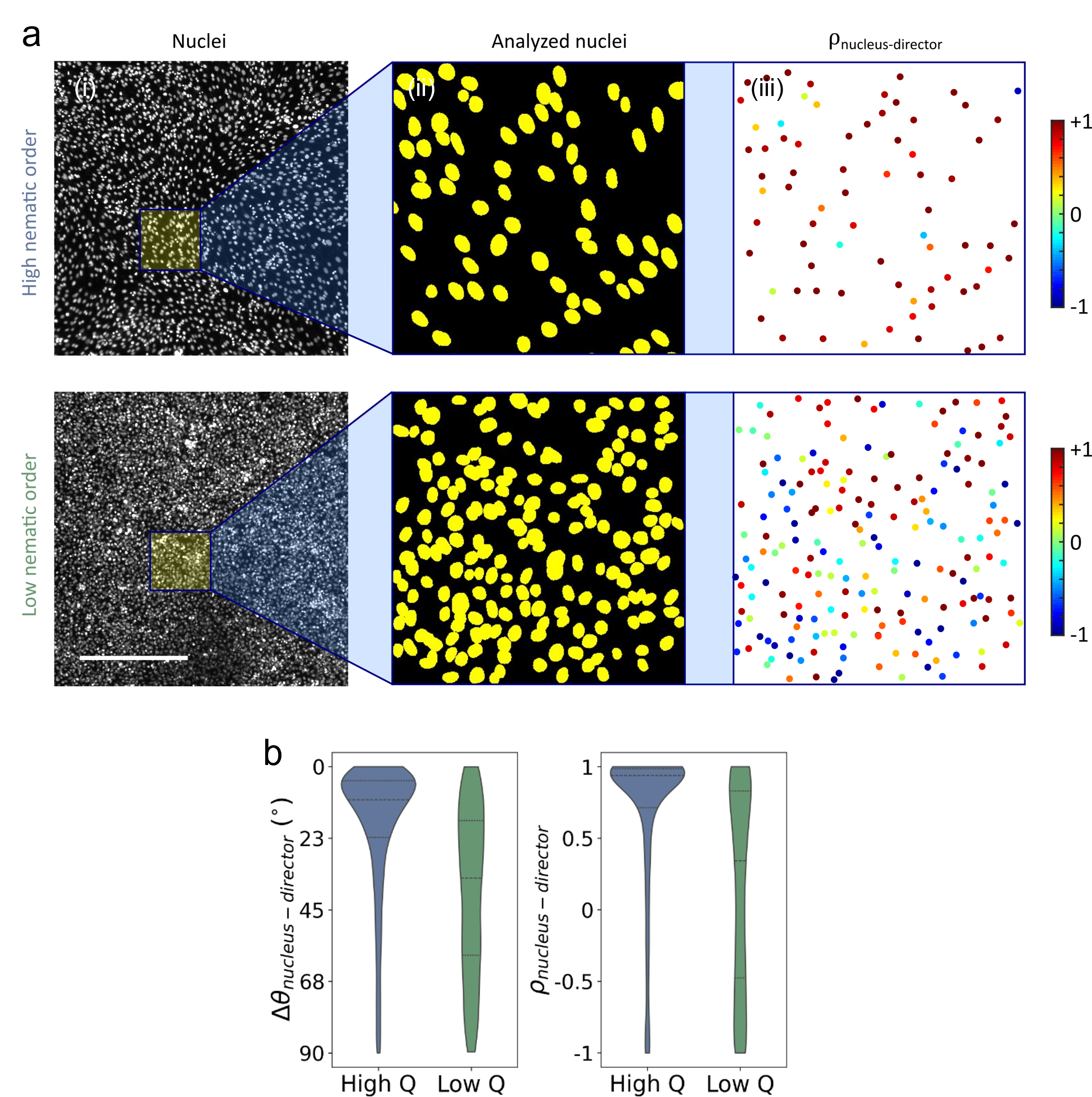}
\caption{\textbf{Correlation between nuclear orientation and local director field.}
\textbf{a}. Representative images of nuclei in regions with high and low nematic order. 
\textbf{b}. Enlarged view of the boxed regions in (a). 
\textbf{c}. Corresponding colormaps of \(\rho\), where each nucleus is colored by its local correlation value 
\(\rho = \cos(2\Delta\theta)\), with \(\Delta\theta\) denoting the angular difference between the nucleus orientation and the local director field obtained from OrientationJ. 
\textbf{d}. Quantitative distributions of \(\rho\) and relative orientation angle \(\Delta\theta\), demonstrating stronger alignment between nuclear and director orientations in the high-order region compared to the low-order region. 
Scale bar: \(500~\mu\mathrm{m}\).}
\label{figS:cell_vs_director}
\end{figure}
\clearpage

\begin{figure}[htbp]
\centering
\includegraphics[width=0.8\textwidth]{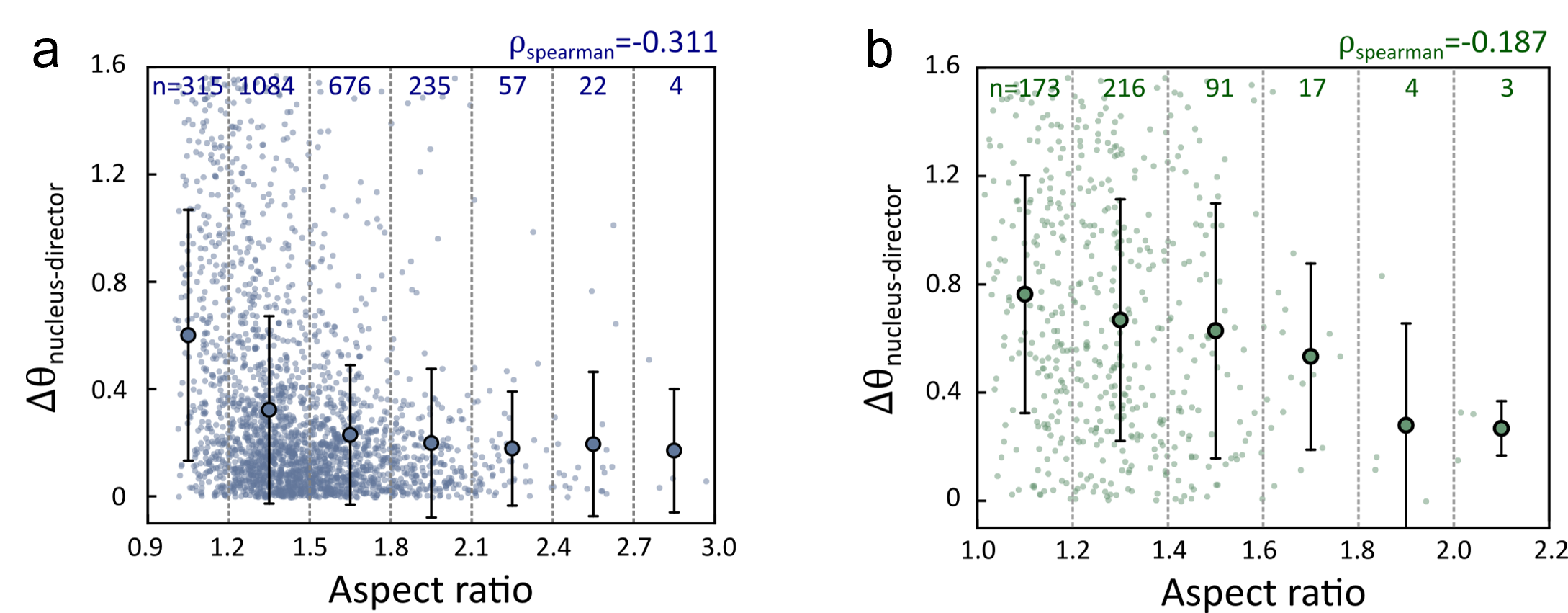}
\caption{\textbf{Dependence of orientation correlation on nuclear morphology.}
Scatter plots showing the relationship between nuclear aspect ratio (AR) and the angular deviation \(\Delta\theta\) between the nucleus orientation and the local director field obtained from OrientationJ for (\textbf{a}) ordered and (\textbf{b}) disordered monolayers. 
Data were binned by AR, and the mean \(\Delta\theta\) within each bin (black circles) was computed to visualize the trend. 
Both datasets exhibit a negative correlation between AR and \(\Delta\theta\), indicating that elongated nuclei (higher AR) tend to align more closely with the local director field. 
The monotonic relationship was quantified using the Spearman correlation coefficient, yielding \(\rho_{\mathrm{spearman}} = -0.311\) for the ordered state and \(\rho_{\mathrm{spearman}} = -0.187\) for the disordered state. 
These results confirm that cells with higher aspect ratios are more accurately represented by OrientationJ as the local orientation of the tissue.}
\label{figS:nucleus_director_AR}
\end{figure}
\clearpage

\begin{figure}[htbp]
\centering
\includegraphics[width=0.8\textwidth]{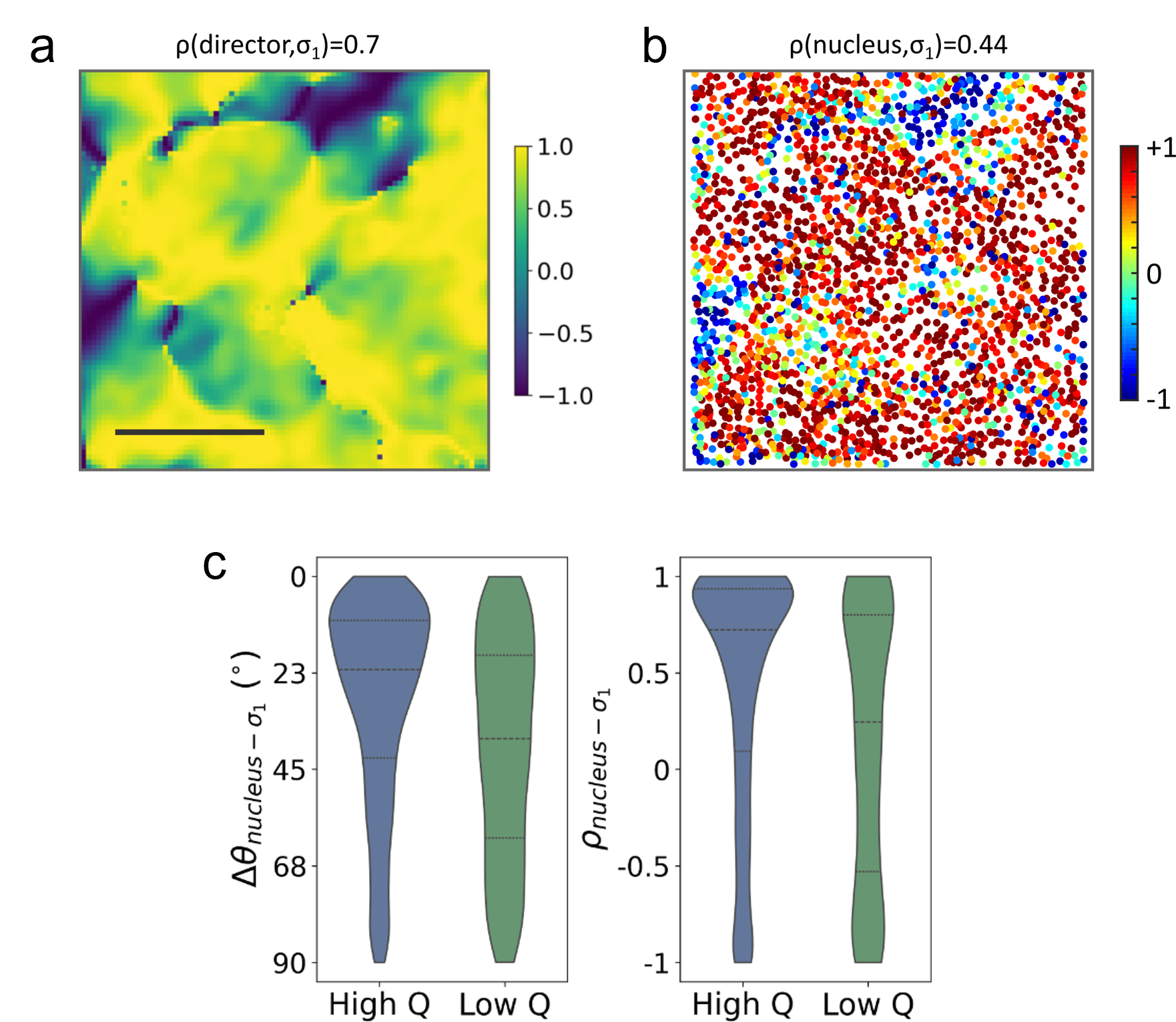}
\caption{\textbf{Correlation between nuclear orientation and principal stress direction.}
\textbf{a}. Spatial map of the local correlation coefficient \(\rho(\text{director}, \sigma_1)\) between the OrientationJ-derived director field and the orientation of the maximum principal stress \(\sigma_1\), as shown in Fig.~4c of the main text. 
\textbf{b}. Corresponding map of \(\rho(\text{nucleus}, \sigma_1)\), directly computed between the nucleus orientation and \(\sigma_1\) orientation for the same ordered monolayer. 
Regions of high and low correlation in (a) and (b) show similar spatial patterns, indicating that the nucleus orientation reliably reflects the cellular alignment relative to the stress field. 
The average correlation between nucleus and stress orientations is \(\rho(\text{nucleus}, \sigma_1) = 0.44\), which is consistent with the expected relation \(\rho(\text{nucleus}, \sigma_1) \approx \rho(\text{director}, \sigma_1) \times \rho(\text{nucleus}, \text{director}) = 0.7 \times 0.7\). 
\textbf{c, d}. Violin plots summarizing the distributions of \(\rho(\text{director}, \sigma_1)\) and \(\rho(\text{nucleus}, \sigma_1)\) for ordered and disordered states. 
These results confirm that in ordered monolayers, both the director field and the nuclear orientation are coherently aligned with the local principal stress direction, whereas such coupling diminishes in disordered states. 
Scale bar: \(500~\mu\mathrm{m}\).}
\label{figS:nucleus_sigma1}
\end{figure}
\clearpage

\begin{figure}[htbp]
\centering
\includegraphics[width=0.8\textwidth]{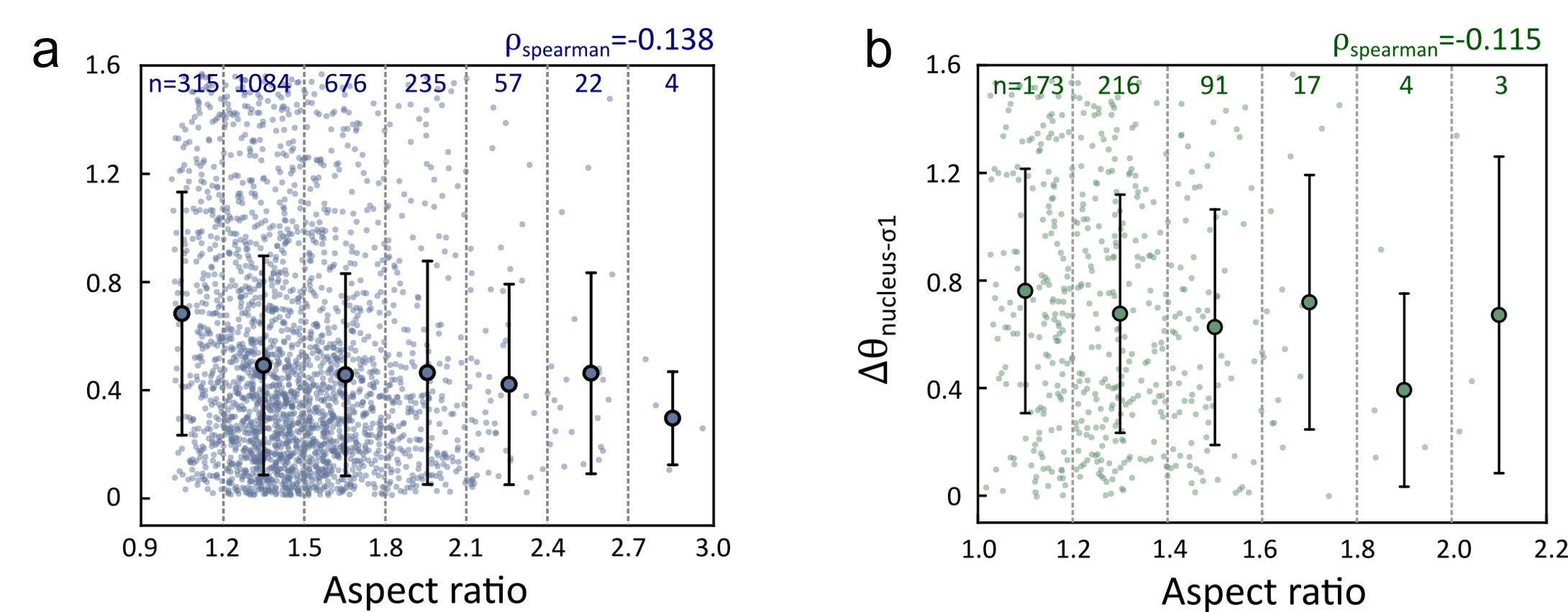}
\caption{\textbf{Dependence of nucleus–stress alignment on nuclear morphology.}
Scatter plots showing the relationship between nuclear aspect ratio (AR) and the angular deviation \(\Delta\theta\) between the nucleus orientation and the maximum principal stress direction \(\sigma_1\) for (\textbf{a}) ordered and (\textbf{b}) disordered cell monolayers. 
Data points were grouped into AR bins, and the mean \(\Delta\theta\) within each bin (black circles) was computed to illustrate the trend. 
Both datasets show a weak but consistent negative correlation between AR and \(\Delta\theta\), indicating that elongated nuclei (higher AR) tend to align more closely with the local \(\sigma_1\) orientation, suggesting that more anisotropic cells are more sensitive to mechanical guidance. 
The correlation strength was quantified by the Spearman correlation coefficient, yielding \(\rho_{\mathrm{spearman}} = -0.138\) for the ordered state and \(\rho_{\mathrm{spearman}} = -0.115\) for the disordered state.}
\label{figS:nucleus_sigma1_AR}
\end{figure}
\clearpage

% \begin{figure}[bth]
%     \centering
%     \includegraphics[width=\linewidth]{Figures/fig-S5-BMSC.jpg}
%     \caption{\textbf{Stress ditribution in different cell types.}
%     \textbf{a}. }
%     \label{fig:S-BMSC}
% \end{figure}
% \clearpage

\begin{figure}[htbp]
\centering
\includegraphics[width=0.9\textwidth]{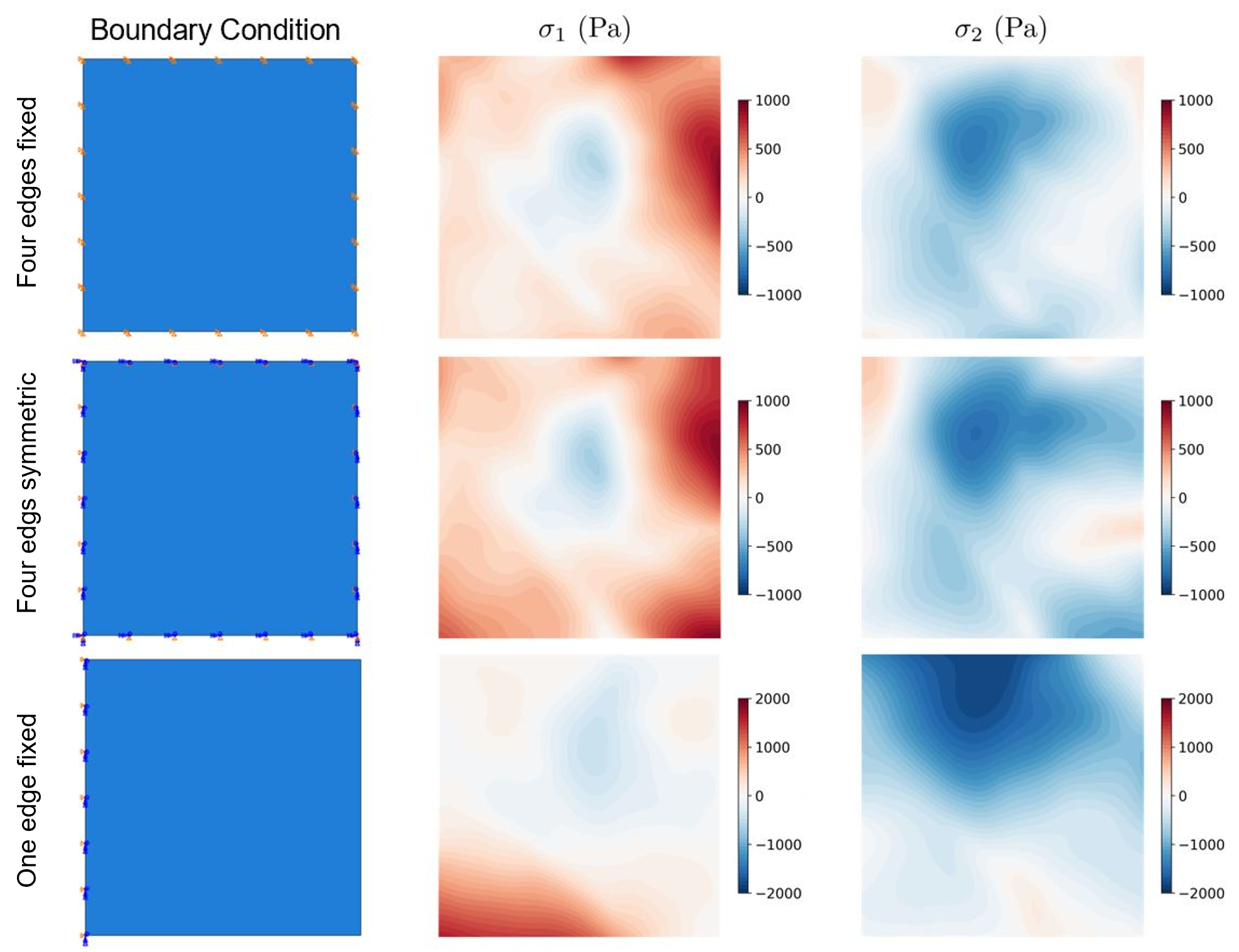}
\caption{\textbf{Effect of boundary conditions on reconstructed stress fields.}
Comparison of the computed principal stress components \(\sigma_1\) and \(\sigma_2\) under different boundary conditions applied in monolayer stress microscopy (MSM). 
From top to bottom: (i) all four edges fixed, (ii) all four edges symmetric (normal displacement constrained to zero), and (iii) one edge fixed and free boundary condition for other three. 
The stress distributions obtained from symmetric and fixed-edge conditions show minimal differences, confirming the robustness of MSM reconstruction against reasonable boundary constraint choices.}
\label{figS:BC_effect}
\end{figure}
\clearpage

\begin{figure}[htbp]
\centering
\includegraphics[width=\textwidth]{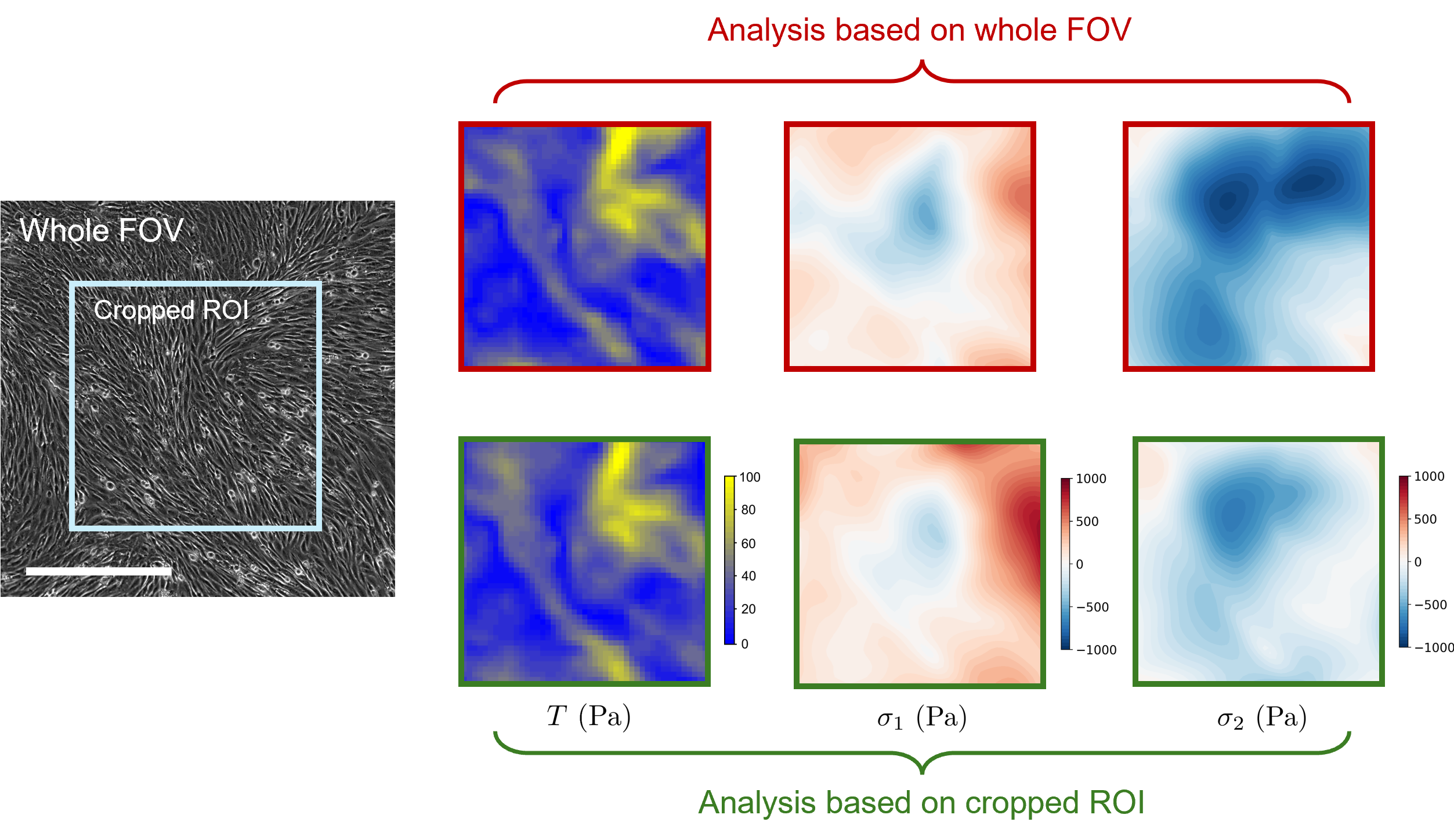}
\caption{\textbf{Validation of field-of-view (FOV) independence in traction and stress reconstruction.} 
To verify that the field size does not affect the computed traction and stress fields, analyses were performed on both the whole field of view (FOV) and a cropped region of interest (ROI). (Left) Representative phase-contrast image showing the whole FOV and the cropped ROI used for comparison. (Top row) Traction and monolayer stress maps obtained by first analyzing the entire FOV and then cropping to the same ROI. (Bottom row) Corresponding maps obtained by first cropping the ROI and then performing the analysis. The two approaches yield nearly identical results, confirming that the traction and reconstructed stress fields are robust against FOV selection, except for small deviations near the boundaries. Scale bar: \(500~\mu\mathrm{m}\).}
\label{figS:FOV}
\end{figure}
\clearpage

\end{document}